%% file: Paper_B1828-11_DataAnalysis.tex
\documentclass[fleqn,usenatbib]{mnras}
\pdfoutput=1
\usepackage[T1]{fontenc}
\usepackage{ae,aecompl}

\usepackage{amsmath}
\usepackage{amssymb}

\usepackage{graphicx}

\usepackage{hhline}

\usepackage{upgreek}
\renewcommand{\pi}{\uppi} 

\newcommand{\tref}{t_{\mathrm{ref}}}
\newcommand{\PATNF}{P^{\mathrm{ATNF}}}

\newcommand{\tauP}{\tau_{\mathrm{P}}}
\newcommand{\tauAge}{\tau_{\mathrm{Age}}}
\newcommand{\PhiO}{\Phi_{\mathrm{obs}}}

\newcommand{\In}{\mathcal{I}}

\newcommand{\nudotOne}{\dot{\nu}_{1}}
\newcommand{\nudotTwo}{\dot{\nu}_{2}}
\newcommand{\Wone}{W_{1}}
\newcommand{\Wtwo}{W_{2}}
\newcommand{\Wave}{\langle W_{10} \rangle}
\newcommand{\tI}[1]{t_{\mathrm{#1}}}
\newcommand{\nudot}{\dot{\nu}}

\newcommand{\M}{\mathcal{M}}
\newcommand{\data}{\textrm{data}}
\newcommand{\params}{\boldsymbol{\theta}}
\newcommand{\yobs}{y^{\textrm{obs}}}
\newcommand{\ydata}{\mathbf{y}^{\mathrm{obs}}}

\newcommand{\mboxcitet}[1]{\mbox{\citet{#1}}}

\newcommand{\bigfigurecaptions}[2]{
$\textbf{A}$: The estimated marginal posterior probability distribution for the
#1 #2 model parameters. $\textbf{B}$: Checking the fit of the model using the
maximum posterior values to the data; see Fig.~\ref{fig: noise-only beam-width
posterior fit} for a complete description.}

\title[Comparing periodic models for PSR B1828-11]{
  Comparing models of the periodic variations in spin-down
       and beam-width for PSR B1828-11}

\author[G. Ashton et al.]{
G.~Ashton,$^{1}$\thanks{E-mail: G.Ashton@soton.ac.uk}
D.I.~Jones,$^{1}$
R.~Prix$^{2}$
\\
$^{1}${Mathematical Sciences and STAG Research Centre,
       University of Southampton,
       Southampton SO17 1BJ} \\
$^{2}${Max Planck Institut f{\"u}r Gravitationsphysik
       (Albert Einstein Institut),
       30161 Hannover, Germany}
}

\pubyear{2015}

\begin{document}
\label{firstpage}
\pagerange{\pageref{firstpage}--\pageref{lastpage}}
\maketitle

\input{Evidence.tex}

\begin{abstract}

We build a framework using tools from Bayesian data analysis to evaluate models
explaining the periodic variations in spin-down and beam-width of PSR B1828-11.
The available data consists of the time averaged spin-down rate, which displays
a distinctive double-peaked modulation, and measurements of the beam-width.
Two concepts exist in the literature that are capable of explaining these
variations; we formulate predictive models from these and quantitatively
compare them.  The first concept is phenomenological and stipulates that the
magnetosphere undergoes periodic switching between two meta-stable states as
first suggested by \citeauthor{Lyne2010}. The second concept, precession, was first
considered as a candidate for the modulation of B1828-11 by \citeauthor{Stairs2000}.
We quantitatively compare models built from these concepts using a Bayesian
odds-ratio.  Because the phenomenological switching model itself was informed
by this data in the first place, it is difficult to specify appropriate
parameter-space priors that can be trusted for an unbiased model comparison.
Therefore we first perform a parameter estimation using the spin-down data, and
then use the resulting posterior distributions as priors for model comparison
on the beam-width data.  We find that a precession model with a simple circular
Gaussian beam geometry fails to appropriately describe the data, while allowing
for a more general beam geometry provides a good fit to the data. The resulting
odds between the precession model (with a general beam geometry) and the
switching model are estimated as
$10^{\oddsBeamwidthModifiedGaussianBeamwidthSwitching\pm\errBeamwidthModifiedGaussianBeamwidthSwitching}$
in favour of the precession model.

\end{abstract}

\begin{keywords}
methods: data analysis --
stars: neutron --
pulsars: individual: PSR B1828-11
\end{keywords}

\section{Introduction}
\label{sec: introduction}

The pulsar B1828-11 demonstrates periodic variability in its pulse timing and
beam shape at harmonically related periods of 250, 500, and 1000 days. The
modulations in the timing was first taken as evidence that the pulsar is
orbited by a system of planets by \mboxcitet{Bailes1993}. A more complete analysis
by \mboxcitet{Stairs2000} concluded that the corresponding changes in the
beam-shape would require at least two of the planets to interact with the
magnetosphere, which does not seem credible. Instead the authors proposed that
the correlation between timing data and beam-shape suggested the pulsar was
undergoing free precession. If true, such a claim would require rethinking of
the vortex-pinning model used to explain the pulsar glitches since the pinning
should lead to much shorter modulation period than observed \citep{Shaham1977},
and fast damping of the modulation \citep{Link2003}.

The idea of precession for B1828-11 has been studied extensively in the
literature: \mboxcitet{Jones2001} derived the observable modulations due to
precession and noted that the electromagnetic spin-down torque will amplify
these modulations.  \mboxcitet{Link2001} fitted a torqued-precession model to the
spin-down and beam-shape followed by \mboxcitet{Akgun2006} where a variety of
shapes and the form of the spin-down torque were tested.  All of these authors
agree that precession is a credible candidate to explain the observed periodic
variations: furthermore to explain the double-peaked spin-down modulations, the
so-called wobble angle must be small while the magnetic dipole must be close to
$\pi/2$.

More recently \mboxcitet{Arzamasskiy2015} updated the previous estimates (based on
a vacuum approximation) to a plasma filled magnetosphere. They also find that
the magnetic dipole and spin-vector must be close to orthogonal, but solutions
could exist where it is the wobble angle which is close to $\pi/2$ while the
magnetic dipole lies close to the angular momentum vector; we will not consider
such a model here, but note it is a valid alternative which deserves testing.

The distinctive spin-down of B1828-11 was analysed by \mboxcitet{Seymour2013} for
evidence of chaotic behaviour. They found evidence that B1828-11 was subject to
three dynamic equations with the spin-down rate being one governing variable.
This further motivates the precession model since it results from applying
Euler's three rigid body equations to a non-spherical body \citep{Landau1969}.

\begin{figure}
\centering
\includegraphics[]{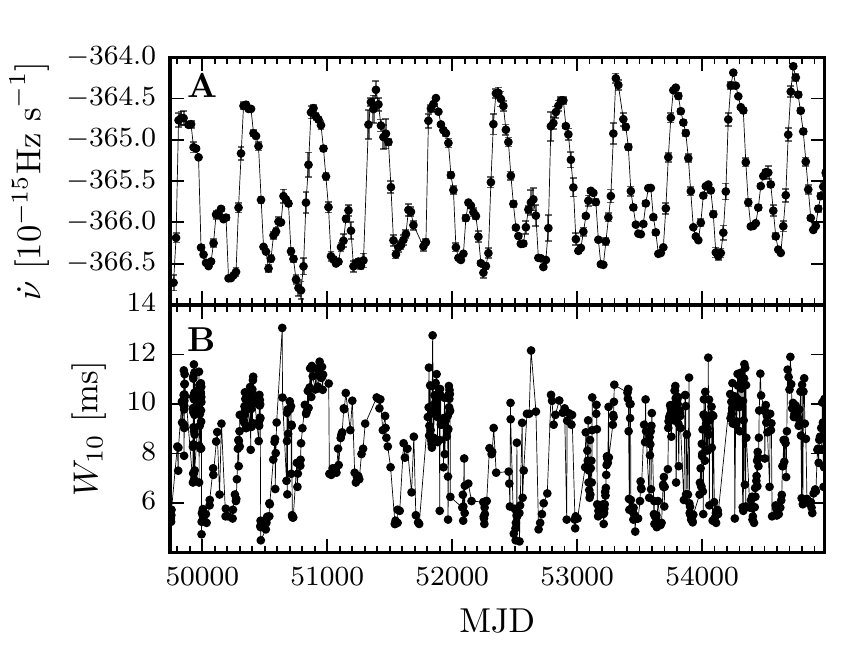}
\caption{Observed data for PSR~B1828-11 spanning from MJD~49710 to MJD~54980.
         In panel $\textbf{A}$ we reproduce the spin-down rate with error-bars
         and in panel $\textbf{B}$ the beam-width $W_{10}$
          (for which no error bars were available). All data
         courtesy of \mboxcitet{Lyne2010}.}
\label{fig: B1828-11 data}
\end{figure}

The precession hypothesis was challenged by \mboxcitet{Lyne2010} when reanalysing
the data.   They noted that in order to measure the spin-down and beam-shape
with any accuracy required time averaging over periods~$\sim100$~days,
smoothing out any behaviour acting on this time-scale. Motivated by the
intermittent pulsar B1931+24, they put forward the phenomenological hypothesis
that instead the magnetosphere is undergoing periodic switching between (at
least) two metastable states. Such switching would result in correlated changes
in the beam-width and spin-down rate. They returned to the data and instead of
studying a time-averaged beam-shape-parameter as done by \mboxcitet{Stairs2000},
they instead considered the beam-width at 10\% of the observed maximum
$W_{10}$. This quantity is time-averaged, but only for each observation
lasting~$\sim1$hr. This makes $W_{10}$ insensitive to any changes which occur on
time-scales shorter than an hour. If the meta-stable states last longer than
this, $W_{10}$ will be able to resolve the switching. The  relevant data was
kindly supplied to us courtesy of \mboxcitet{Lyne2010}, and is reproduced in
Fig.~\ref{fig: B1828-11 data}. From these observations,
\mboxcitet{Lyne2010} concluded that the individual measurements of $W_{10}$ for
B1828-11 did in fact appear to switch between distinct high and low values, as
opposed to a smooth modulation between the values, with this switching
coinciding with the periodic changes in the spin-down. On this basis, they
interpret the modulations of B1828-11 as evidence it is undergoing periodic
switching between two magnetospheric states. When studying another pulsar which
also displays double-peaked spin-down modulations, \citet{Perera2015} extended
the switching model, as discussed in Sec.~\ref{sec: switching}, to be capable
of producing the double-peak spin-down rate; it is this modification of the
switching model which we will be comparing with precession.

In our view, it is not immediately clear by eye whether the data presented in
Fig.~\ref{fig: B1828-11 data} is sufficient to rule out or even favour either
of the precession or switching interpretations.  For this reason, in this work
we develop a framework in which to evaluate models built from these concepts
and argue their merits quantitatively using a Bayesian model comparison. We
note that a distinction must be made between a conceptual idea, such as
precession, and a particular predictive model built from it. As we will see,
each concept can generate multiple models, and furthermore we could imagine
using a combination of precession and switching,  with the precession acting as
the `clock' that modulates the probability of the magnetosphere being in one
state or the other, an idea developed by  \mboxcitet{Jones2012}. The models
considered here cover the precession and switching interpretations, but we do
not claim the models to be the `best' that these hypotheses could produce.

The rest of the work is organised as follows: in Sec.~\ref{sec: methodology} we
will describe the framework to fit and evaluate a given model, in
Sec.~\ref{sec: defining and fitting the models} we will define and fit several
predictive models from the conceptual ideas, and then in Sec.~\ref{sec:
estimating the odds-ratio} we shall tabulate the results of the model
comparison. Finally, the results are discussed in Sec.~\ref{sec: discussion}.

\section{ Bayesian Methodology}
\label{sec: methodology}
We now introduce a general methodology to compare and evaluate models for this
form of data. The technique is well practised in this and other fields and so
in this section we intend only to give a brief overview; for a more complete
introduction to this subject see \cite{jaynes2003probability,
gelman2013bayesian, sivia1996data}.

\subsection{The odds-ratio and posterior probabilities}

There are two issues that we wish to address.  Firstly, given two models, how
can one say which is preferred, and by what margin?  Secondly, assuming a given
model, what can be said of the probability distribution of the
parameters that appear in that model?

We can address the first issue by making use of Bayes theorem for the
probability of model $\M_{i}$ given some data:
\begin{equation}
    P(\M_{i} | \data) = P(\data | \M_{i}) \frac{P(\M_{i})}{P(\data)}.
\label{eqn: bayes theorem}
\end{equation}
The quantity $P(\data | \M_{i})$ in known as the \emph{marginal likelihood} of
model $\M_i$ given the data.

In general we cannot compute the probability given in equation (\ref{eqn: bayes
theorem})  because we do not have an exhaustive set of models to calculate
$P(\data)$.  However, we can compare two models, say $A$ and $B$, by
calculation of their \emph{odds-ratio}:
\begin{equation}
\mathcal{O} = \frac{P(\M_{A}|\data)}{P\M_{B}|(\data)}=
            \frac{P(\data|\M_{A})}{P(\data|\M_{B})}
            \frac{P(\M_{A})}{P(\M_{B})}.
\label{eqn: odds-ratio}
\end{equation}
In the rightmost expression, the first factor is the ratio of the marginal
likelihoods (also known as the \emph{Bayes factor}) which we will discuss
shortly, while the final factor reflects our prior belief in the two models. If
no strong preference exists for one over the other, we may take a
non-informative approach and set this equal to unity.  We will follow this
approach in what follows below.

We need to find a way of computing the marginal likelihoods, $P(\data |
\M_{i})$.  To this end, consider a single model  $\M_{i}$ with model parameters
$\params$, and define $P(\data| \params, \M_{i})$ as the \emph{likelihood
function} and $P(\params | \M_{i})$ as the \emph{prior distribution} for the
model parameters.   We can then perform the necessary calculations by making
use of
\begin{equation}
    P(\data|\M_{i}) = \int
                      P(\data| \params, \M_{i})P(\params | \M_{i})
                      d\params.
\label{eqn: marginal likelihood}
\end{equation}

The likelihood function can also be used to explore the second issue of
interest, by calculating the \emph{joint-probability distribution} for the
model parameters, also known as the \emph{posterior probability distribution}:
\begin{align}
P(\params| \data, \M_{i}) =
\frac{P(\data| \params, \M_{i})P(\params | \M_{i})}
{P(\data| \M_{i})}.
\end{align}
Note that the marginal likelihood $P(\data| \M_{i})$ described above plays the
role of a normalising factor in this equation.

In general the integrand of Eqn.~\eqref{eqn: marginal likelihood} makes
analytic, or even simple numeric integration difficult or impossible. This is
the case for the probability model that we will use and so instead we must turn
to sophisticated numerical methods.  For this study we use Markov-Chain
Monte-Carlo (MCMC) techniques which simulate the joint-posterior distribution
for the model parameters up to the normalising constant
\begin{align}
P(\params | \data, \M_{i}) \propto
P(\data| \params, \M_{i})P(\params | \M_{i}).
\end{align}
In particular we will use the \mboxcitet{Foreman-Mackay2013} implementation of the
affine-invariant MCMC sampler \citep{Goodman2010} to approximate the posterior
density of the model parameters.  Further details of our MCMC calculations can
be found in appendix~\ref{sec: procedure for the mcmc parameter estimation}.

Once we are satisfied that we have a good approximation for the joint-posterior
density of the model parameters we discuss how to recover the normalising
constant to calculate the odds-ratio in Sec.~\ref{sec: estimating the
odds-ratio}.

\subsection{Signals in noise}
We now need to build a statistical model to relate physical models for the
spin-down and beam-width to the data observed in Fig.~\ref{fig: B1828-11 data}.
To do this we will turn to a method widely used to search for deterministic
signals in noise.

We assume our observed data $\yobs$ is a sum of a stationary zero-mean
Gaussian noise process
$n(t, \sigma)$ (here $\sigma$ is the  standard deviation of the noise process)
and a signal model $f(t| \M_{j}, \params)$ (where $\params$ is a vector of the
model parameters) such that
\begin{equation}
\yobs(t_i| \M_{j}, \params, \sigma) = f(t_i|\M_{j}, \params) + n(t_i, \sigma).
\label{eqn: yobs def}
\end{equation}

Given a particular signal model, subtracting the model from the data
should, if the model and model parameters are correct, leave behind a Gaussian
distributed residual - the noise. That is
\begin{equation}
\yobs(t_i| \M_{j}, \params, \sigma) - f(t_i|\M_{j}, \params) \sim N(0, \sigma).
\end{equation}

The data, for either the spin-down or beam-width, consists of $N$ observations
$(\yobs_{i}, t_{i})$. For a
single one of these observations,
the probability distribution given the model and model parameters is
\begin{align}
P(y_{i}^{\textrm{obs}}| \M_j, \params, \sigma) =
\frac{1}{\sigma\sqrt{2\pi}}
\exp\left\{\frac{-\left(f(t_{i}|\M_j, \params)
            - y_{i}\right)^{2}}{2\sigma^{2}}\right\}.
\end{align}
The likelihood is the product of the $N$ probabilities
\begin{equation}
P(\ydata| \M_{j}, \params, \sigma) =
              \prod_{i=1}^{N} P(y_{i}^{\textrm{obs}}|\M_{j}, \params, \sigma),
\label{eqn: signals in noise likelihood}
\end{equation}
where $\ydata$ denotes the vector of all the observed data.

In Sec.~\ref{sec: defining and fitting the models} we will define the physical
models, $f(t| \M_j, \params)$, for the precession and
switching interpretations; for now we recognise that once defined, we may
calculate the likelihood of the data under the model using Eqn.~\eqref{eqn:
signals in noise likelihood}.

\subsection{Choosing prior distributions}
\label{sec: setting prior distributions}

In the previous section we have developed the likelihood function $P(\data|
\params, \M_{j})$ for any arbitrary model producing a deterministic signal
$f(t_{i}|\M_j, \params)$ in noise. To compare between particular models, using
Eqn.~\eqref{eqn: odds-ratio}, we must compute the marginal likelihood as
defined in Eqn.~\eqref{eqn: marginal likelihood} which requires a prior
distribution $P(\params|\M_{j})$.

The choice of prior distribution is important in a model comparison since it
can potentially have a large impact on the resulting odds-ratio.  In general we
want to use astrophysically informed priors wherever possible, or suitable
uninformative (but proper) priors otherwise. However, the switching model
presents a particular challenge in this respect, as its switching parameters
(cf.\ Sec.~\ref{sec: switching}) are ad-hoc and purely phenomenological, and
were initially informed by the same data we are trying to test the models on.
It is therefore important to avoid potential circularity in properly assessing
the prior volume of its parameter space, which affects the relevant ``Occam
factor'' for this model (e.g.\ see \mboxcitet{mackay2003information}).

To resolve this, we will make use of the availability of two different and
independent data sets: the spin-down and the beam-width data.  First, we will
perform parameter estimation using the spin-down data with astrophysical priors
where possible and uniform priors based on crude estimates from the data
otherwise. For the model parameters common to both the spin-down and
beam-width models, we will use the posterior distributions from the spin-down
data as prior distributions for the beam-width model. For the remaining
beam-width parameters which are not common to both the spin-down and beam-width
models we will use astrophysically-motivated priors. In this way we can do
model comparisons based on the beam-width data using proper, physically
motivated priors. In addition, this enforces consistency between the beam-width
and spin-down solutions: for example constraining the two to be in phase.

An obvious alternative is to do the reverse and use the beam-width data to
determine priors for the spin-down data. However, for both models, we found
difficulties in obtaining good quality posteriors when conditioning on the
beam-width data with uniform priors based on crude estimates.
Specifically, we found the posteriors to be non-Gaussian and multimodal.
To deal with this we would
need to use a more sophisticated methodology than that discussed in
appendix~\ref{sec: procedure for the mcmc parameter estimation}. By contrast this
is not the case when conditioning on the spin-down data first (results
presented in Sec.~\ref{sec: defining and fitting the models}). This is expected
since, even by eye, we see that the spin-down data contains an easily visible
`signal', while the beam-width data is relatively `noisy'.  For this work we
are primarily interested in laying out the framework to perform model
comparisons and either method should suffice and give the same solution.  For
now then, we will use the more straight-forward method of using the spin-down
data to set priors for the beam-width.

\section{Defining and fitting the models}
\label{sec: defining and fitting the models}
In this section we will take each conceptual idea (precession or switching) and
define a predictive signal model $f(t|\M_j, \params)$. Each concept may motivate
multiple signal models: already we have seen the extension to the original
\mboxcitet{Lyne2010} switching model by \mboxcitet{Perera2015}. In this work we do not
aim to exhaust all known models and are well aware that more models exist that
have not yet been considered.

For each concept, we will first discuss the theoretical model, then discuss the
choice of priors and finally the resulting posterior and posterior-predictive
checks. For both these concepts we build models for both the spin-down and
beam-width using the former to inform the priors for the latter as described in
Sec.~\ref{sec: setting prior distributions}. Model comparisons will be made on
the beam-width data only. In addition to these two concepts, we will also
consider a noise-only model for the beam-width data.

It is worth stating that by using the signals-in-noise statistical model, we do
not make any assumptions on the cause of the noise other than requiring it to
be stationary and Gaussian (cf.\ \mboxcitet{jaynes2003probability}). Given the
uncertain physics of neutron stars and the measurement of pulses, it seems
likely the noise will contain contributions both from the neutron star itself,
and from the measurement process, with the former dominating. We will add a
subscript to the noise component $\sigma_{[\dot{\nu}, W_{10}]}$ to distinguish
between the two data sources.

\subsection{Noise-only model}
\label{sec: noise-only}
\subsubsection{Defining the noise-only beam-width model}

Before evaluating the precession and switching hypothesis, let us first
consider a noise-only model.  This will introduce some generic concepts and
provide a benchmark against which to test other models. The noise-only model
asserts that the beam-width data (as seen in panel B of Fig.\ref{fig: B1828-11
data}) does not contain any periodic modulation, but is the result of noise
about a fixed beam-width: the signal model $f(t) = W_{10}$ is a constant.

We will not consider the spin-down data under such a hypothesis since it is the
beam-width data alone that we will use to make model comparisons and it is
clear by eye that such a model is incorrect.

\subsubsection{Fitting the model to the beam-width data}
\label{sec: noise-only fitting the model}

For the noise-only model we have two parameters which require a prior: the
constant beam-width $W_{10}$ and the noise $\sigma_{W_{10}}$. For the
beam-width we will set a prior using astrophysical data on the period $P$
from the ATNF database
(available at \url{www.atnf.csiro.au/people/pulsar/psrcat}, for a description
see \cite{ATNF}). This value, $\PATNF=0.405043321630\pm 1.2\times10^{-11}$~s,
provides a strict upper bound on $W_{10}$, although typically integrated
pulse profiles only occupy between 2\% and 10\% of the period \citep{Lyne1988}.
Therefore we will use a uniform prior on $[0, 0.1 \PATNF]$ for $\Wave$.
The choice of 10\% adds a degree of ambiguity into the model comparison since
varying it will change the odds-ratio; we investigate this in
Sec.~\ref{sec: effect of the choice of prior}.

For the noise parameter
$\sigma_{W_{10}}$ we will use a prior $\textrm{Unif}(0, 5)$~ms
based on a crude estimate from the data. We must be careful here as by doing
this we are in a sense using the data twice, but this will not introduce bias
into the model comparison provided the same prior is applied for all beam-width
models.

\begin{figure}
\centering
\includegraphics[]{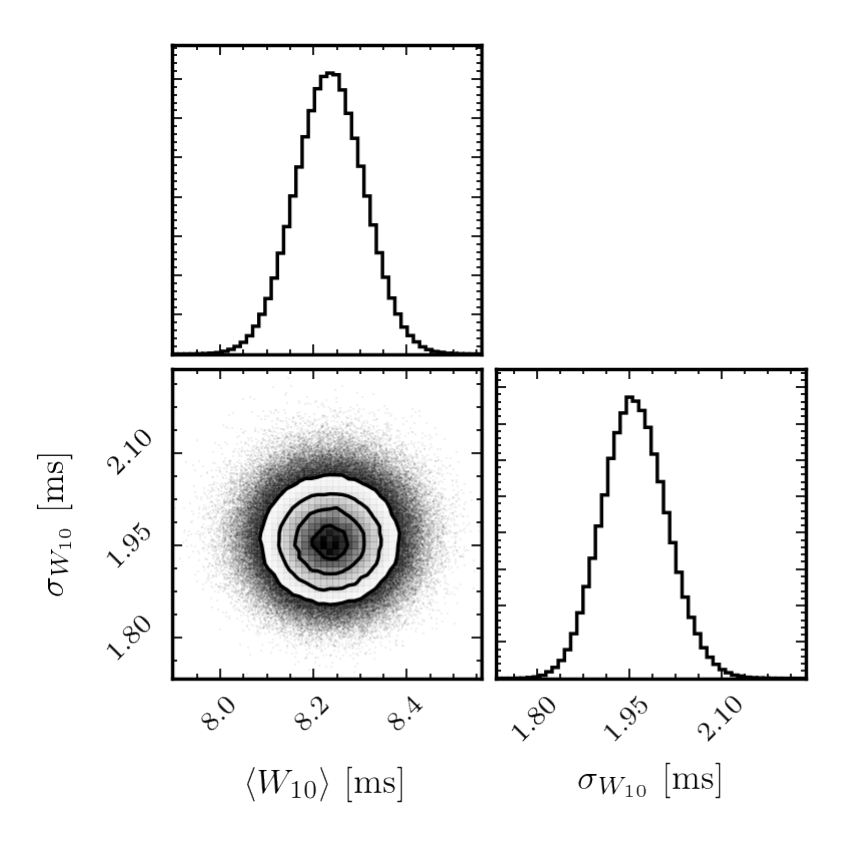}
\caption{The estimated marginal posterior probability distributions for the
noise-only model parameters of the beam-width data.}
\label{fig: noise-only beam-width posterior}
\end{figure}

The MCMC simulations converge quickly to a normal distribution as shown in
Fig.~\ref{fig: noise-only beam-width posterior}. Of note is the mode of
$\sigma_{W_{10}}\sim 2$~ms; this is the Gaussian noise required to explain the
variations in $W_{10}$ about a fixed mean. For other models, we hope to explain
some of the variations with periodic modulation and the rest with Gaussian
noise. So for these models we should expect $\sigma_{W_{10}} < 2$~ms.

\begin{figure}
\centering
\includegraphics[width=0.5\textwidth]{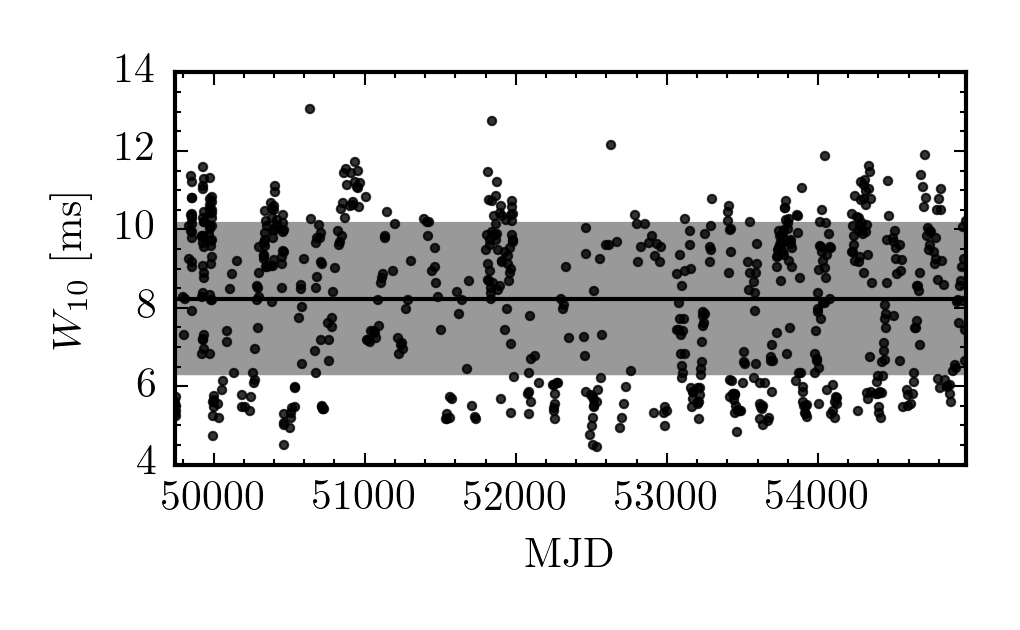}
\caption{Posterior predictive check of the fit of the noise-only model
posterior distribution to the data: the solid black line is the maximum
posterior estimate (MPE), i.e. the model prediction when the parameters are set
equal to the values corresponding to the peaks of the posterior probability
distributions.   The shaded region indicates the MPE of the 1-$\sigma_{W_{10}}$
noise about the beam-width model, black dots are the original data.}
\label{fig: noise-only beam-width posterior fit}
\end{figure}

In Fig.~\ref{fig: noise-only beam-width posterior fit} we plot the
\emph{maximum posterior estimate} (MPE) of the signal alongside the data, i.e.
the model prediction when the parameters are set equal to the peak values of
the posterior probability distributions.  This figure demonstrates that, for
the noise-only model, the observed $W_{10}$ has a mean value of approximately
$8$~ms, then all the variations about this mean are due to the noise. In the
following section we will develop models where at least some the variation is
explained by periodic modulations.

\subsection{Switching model}
\label{sec: switching}

The switching idea is phenomenological and we will build the model based on the
modification of \mboxcitet{Lyne2010} by \mboxcitet{Perera2015}: that is we assume the
magnetosphere switches between two meta-stable states \emph{twice} during a
single period (the motivation for this is discussed in Sec.~\ref{sec: spin-down
rate}). For this work we will assume the switching to be deterministic,
although improvements could be made by allowing the switching time to dither,
or probabilistic variations in the switching states themselves; see
\mboxcitet{Lyne2010} for some exploration of such ideas. This fully deterministic
model captures the primary features without explaining the underlying physics,
for example the cause of the switching.  Both \mboxcitet{Jones2012} and
\mboxcitet{Cordes2013} have worked to improve the physical motivations for the
switching and provide a consistent picture. Nevertheless, in this work we
choose to use the simple phenomenological model as a basis, which can be
improved upon in future work.

\subsubsection{Defining the spin-down rate model}
\label{sec: spin-down rate}

The model proposed by \mboxcitet{Lyne2010} poses two states for the magnetosphere
which we will label as $S_{1}$ and $S_{2}$. Then associated to each of these
states is a corresponding spin-down rate $\nudotOne$ and $\nudotTwo$. The
smoothly varying spin-down that we observe is a result of the time-averaging
process required to measure the spin-down rate. \mboxcitet{Lyne2010} suggested a
square-wave-like switching with a duty rate measuring the fraction of time
spent in one state compared to the other. They also proposed a dither in the
switching period which will obscure the periodicity, and may give rise to
low-frequency structure; we will not consider the dither in this work, but will
investigate it in future work. While studying PSR~B0919+06, which also
demonstrates a double-peaked spin-down rate like B1828-11, the authors of
\mboxcitet{Perera2015} realised that a (deterministic) switching model which flips
once per cycle is incapable of explaining the double-peak observed in the
spin-down rate (in particular that one peak is systematically smaller than the
other). In order to explain this double-peaked structure, they propose that the
mode-changes responsible for switching in the spin-down rate must be
doubly-periodic: that is the spin-down rate changes state \emph{twice} during a
single cycle. Other modifications, such as introducing a third magnetospheric
state, are possible, but in this work we will apply the \mboxcitet{Perera2015}
switching model to B1828-11.

\begin{figure}
\centering \includegraphics[width=0.5\textwidth]{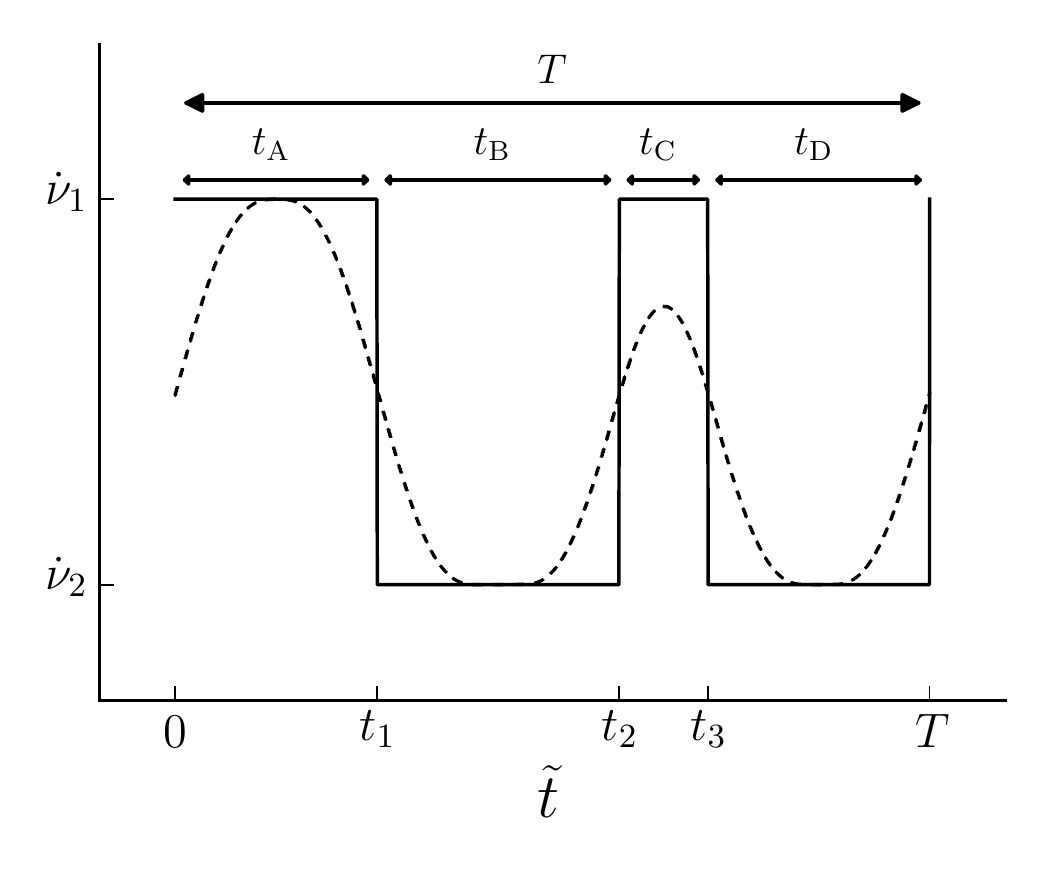}
\caption{Schematic of the doubly-periodic spin-down rate model proposed by
\mboxcitet{Perera2015}. The solid line is the underlying spin-down evolution while
the dashed line indicates the measured time-averaged quantity. In this
instance, the time-average window is longer than $\tI{C}$, but shorter than the
other three durations.}
\label{fig: perera illustration}
\end{figure}

We now discuss the particular formulation of this model used in this
analysis, firstly defining the \emph{underlying} spin-down model and then the
time-averaging process. To aid in this discussion we plot both the underlying
spin-down model and time-average in Fig.~\ref{fig: perera illustration} and
gradually introduce each feature.

We begin by defining $\tilde{t} = (t-\tref) + \phi_{0}T\mod(T)$ where $\phi_0 \in [0,
1]$ is an arbitrary phase offset and $\tref$ is a reference time. For all models
in this study we will set $\tref=$~MJD~49621 to coincide with the epoch at which
the ATNF database \mboxcitet{Manchester1977} records measurements for B1828-11.
Then the function which generates the switching is
\begin{align}
\dot{\nu}_{\mathrm{u}}(t) = \left\{
    \begin{array}{ccc}
    \nudotOne & \mathrm{if} & 0 < \tilde{t} < t_{1}
    \textrm{ or } t_{2} < \tilde{t} < t_{3} \\
    \nudotTwo & \mathrm{if} & t_{1} < \tilde{t} < t_{2}
    \textrm{ or } t_{3} < \tilde{t} < T\\
    \end{array}
    \right. ,
\end{align}
where the subscript `u' denotes that this is the underlying spin-down model.

There are multiple ways to parametrize the switching times in the model. For
the data analysis, we have chosen to parametrize by the total cycle duration
$T$ and three of the segment durations $\tI{A}$, $\tI{B}$ and $\tI{C}$.

This model is subject to label-switching degeneracy in the choice of
$\nudotOne$ and $\nudotTwo$ and also between the various time-scales and
initial phase $\phi_0$.  This degeneracy may cause difficulties in the MCMC
search algorithm, and we therefore fix this gauge freedom by specifying that
$|\nudotOne| < |\nudotTwo|$, $\tI{A} \ge \tI{C}$, and we require that $\tI{A}$
refers to a segment where $\dot{\nu} = \nudotOne$.

Based on a cursory inspection of the observed B1828-11 spin-down rate (see
Fig.~\ref{fig: B1828-11 data} panel A), it is clear that the secular
second-order spin-down rate is non-zero. To model this we will include a
constant $\ddot{\nu}$ in the underlying spin-down model
\begin{align}
    \nudot(t) = \dot{\nu}_{\mathrm{u}}(t) + \ddot{\nu}(t-\tref)\,.
\label{eqn: switching spin-down model}
\end{align}
This gives the intrinsic spin-down rate of the pulsar which would be observed if
measurements could be taken \emph{without} time-averaging

To simulate the observed spin-down rate, we could time-average Eqn.~\eqref{eqn:
switching spin-down model} directly. Instead we choose to mimic the data
collection process responsible for the time-averaging. Let
us first discuss the data collection process as described by \mboxcitet{Lyne2010}.
Observers start with the time-of-arrival of pulsations, which is a measure of
the pulsar rotational phase. Taking a 100~day window of data, starting at the
earliest observation, a second-order Taylor expansion in the phase is fitted to
the data yielding a measurement of $\dot{\nu}$. Then the processes is repeated,
sliding the window in intervals of $25$ days over the whole data set. The
measured $\dot{\nu}$ values at the centre of each window gives a time-averaged
spin-down rate.

To mimic this data collection process, we first integrate Eqn.~\eqref{eqn:
switching spin-down model} twice to generate the phase and then repeat the
above process.  When integrating, we can ignore the arbitrary phase and
frequency offsets since we discard them when calculating the spin-down rate.
The resulting spin-down rates constitutes our signal model which is the
time-average of Eqn.~\eqref{eqn: switching spin-down model}.  A schematic
representation of the sort of spin-down that is then found is given by the
dotted curve in Figure  Fig.~\ref{fig: perera illustration}.  Clearly, the
time-averaged spin-down is much smoother than than the underlying spin-down.

It is worth taking a moment to realise that the relation of the time-average
spin-down to the underlying model $\nudot$ depends on both the segment
durations and the length of the time average ($\tI{ave}=100$~days). For the
$i^{th}$ segment, if the duration $t_{i} > \tI{ave}$ then the time-averaged
spin-down will ``saturate'' and have a flat spot as in segment A of the
illustration in Fig.~\ref{fig: perera illustration}. On the other hand, if
$t_{i} < \tI{ave}$ then the maximum spin-down rate in this segment will be a
weighted sum of the two underlying spin-down rates as in segment B of the
illustration. The weighting is determined by the amount of time the underlying
spin-down rate spends in each state during the time-average window.

\subsubsection{Parameter estimation for the spin-down}

In Table~\ref{tab: perera spin-down prior} we list the selected priors. For
$\ddot{\nu}$ we define
$\ddot{\nu}^{\tiny\mathrm{ATNF}}=(8.75\pm0.09)\times10^{-25}$~s$^{-3}$ (the
value from the ATNF catalogue) and use a normal prior with this mean and
standard deviation. In the tables we show the difference with respect to this
value. For the remaining parameters we select uniform priors using
crude estimates of the data in panel A of Fig.~\ref{fig: B1828-11 data}.
As previously mentioned, this means we are using the
data twice: once in setting up the priors and once for the fitting. Therefore
it would be inappropriate to use the results in a model comparison and this is
not our intention: we want to use the posterior distribution as a prior for the
beam-width parameter estimation.
\begin{table}
\centering
\caption{Prior distributions for the spin-down switching model.}
\label{tab: perera spin-down prior}
\input{Table_Spindown_Switching_prior.tex}
\end{table}

\begin{figure*}
\centering
\includegraphics[]{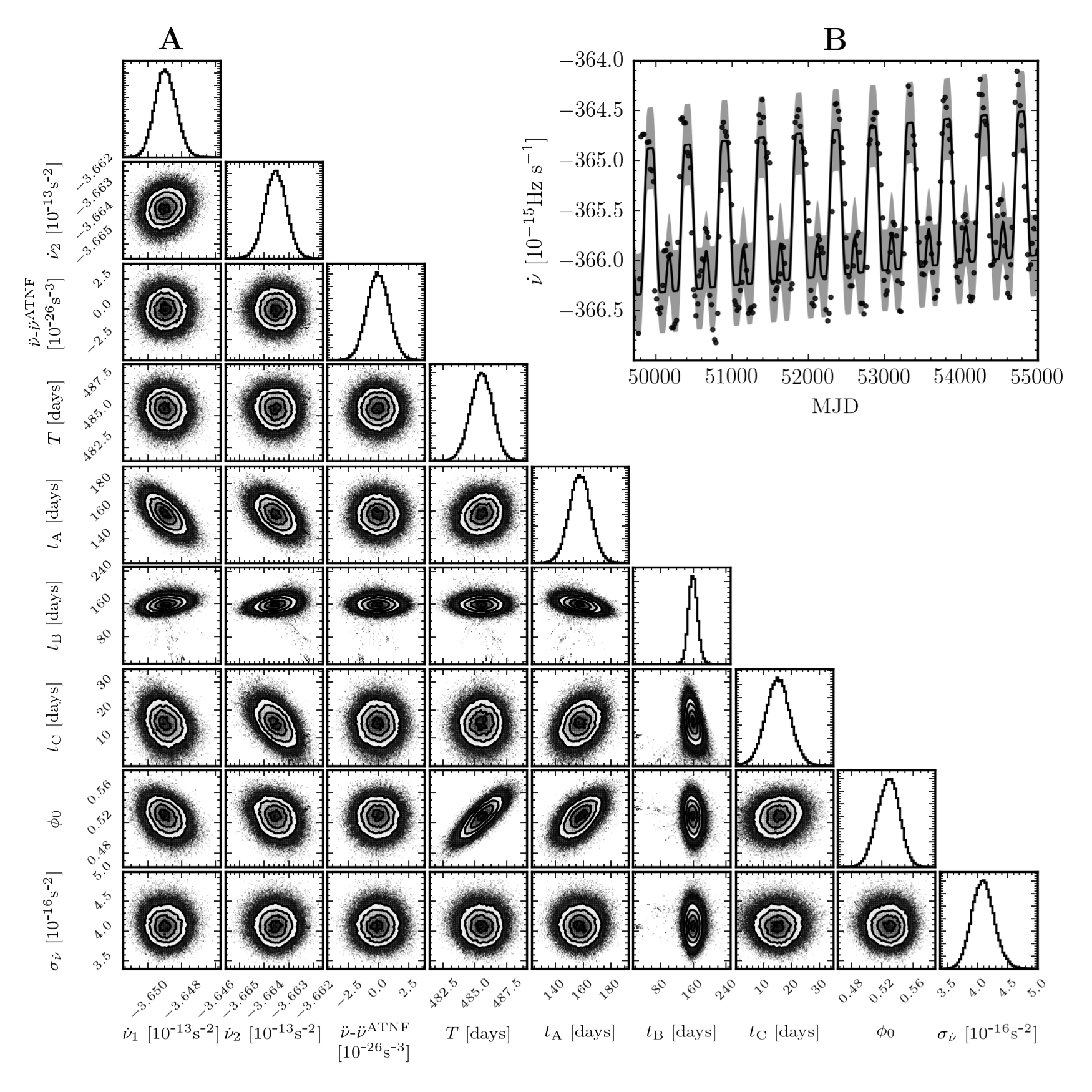}
\caption{\bigfigurecaptions{Switching}{spin-down}}
\label{fig: perera spin-down posterior}
\end{figure*}

For the spin-down data under the switching model, the MCMC simulations converge
quickly to unimodal and approximately Gaussian distributions. The distributions are plotted
in Fig.~\ref{fig: perera spin-down posterior}A, and we summarise the results by
their mean and standard-deviation in Table~\ref{tab: perera spin-down
posterior}. In the case of parameters with uniform priors, this indicates that the data
is informative and a well defined `best-fit' has been selected. For $\ddot{\nu}$
the posterior has not departed significantly from the (informative) prior meaning that the
data agrees with the prior.
\begin{table}
\centering
\caption{Posterior estimates for the spin-down switching model.}
\label{tab: perera spin-down posterior}
\input{Table_Spindown_Switching_posterior.tex}
\end{table}

To check that our fit is sensible, we plot the observed spin-down data in
Fig.~\ref{fig: perera spin-down posterior}B alongside the maximum posterior
estimate for the signal. The relative size of the noise component informs us
how well the model fits the data: if $\sigma_{\dot{\nu}}$ is of a similar size
to the variations in spin-down rate, then the model does poorly and we require
a large noise component. In this case the noise-component is smaller than the
variations in spin-down rate and the signal model explains most of the
variations in the data.

Comparing the maximum posterior values of the four segment times to the
baseline on which we time-average (fixed at 100~days) can give an insight into how the
model has best fit the data. If we take the posterior of $\tI{C}$, we find it
has a mean value of $\sim 15$~days which is significantly shorter than the
baseline on which we time-average. For the other three segments, their durations are
longer than this baseline. The reason that the fit in Fig.~\ref{fig: perera
spin-down posterior}B has one maxima smaller than the other, is because the
segment duration for that segment, $\tI{C}$, is shorter than the time-average
baseline. This is expected and was precisely the motivation for using the model
proposed by \mboxcitet{Perera2015}; a switching model split into only two segments
could not produce this feature.

\subsubsection{Defining the beam-width model}

For the beam-width, we assume that changes in the spin-down rate directly
correlate to changes in this beam-width through changes in the beam geometry.
Since we require the switching to be doubly-periodic for the spin-down to make
sense, so we must require the beam-width to be doubly periodic. That is we
define the beam-width model to be
\begin{align}
W(t) \left\{
    \begin{array}{ccc}
    \Wone & \mathrm{if} & 0 < \tilde{t} < t_{1}
    \textrm{ or } t_{2} < \tilde{t} < t_{3}\\
    \Wtwo & \mathrm{if} & t_{1} < \tilde{t} < t_{2}
    \textrm{ or } t_{3} < \tilde{t} < T\\
    \end{array}
    \right..
\label{eqn: switching beam-width model}
\end{align}

\mboxcitet{Lyne2010} noted that the larger beam-widths tended to correlate with the
lower (absolute) spin-down rate ($\nudotOne$ in our model). We could fix this
by requiring that $\Wone > \Wtwo$ (recalling that we set $|\nudotOne| <
|\nudotTwo|$), but instead we will not implement such a constraint and allow
the data to decide.  As with the spin-down, to break the degeneracy in the
times we will require again that $\tI{A} \ge \tI{C}$. We will not assume any
secular changes in the beam-width for simplicity.

\subsubsection{Parameter estimation for the beam-width}

Having obtained a sensible fit to the spin-down data, we use the resulting
posteriors (as summarised in Table~\ref{tab: perera spin-down posterior}) to
inform our priors for the beam-width data. We can do this only for those
parameters common to both the beam-width and spin-down predictions of the
switching model: namely the four time-scales, and the phase-offset. We would
like to relate the spin-down rates $\nudotOne$ and $\nudotTwo$ to the
beam-widths. However, the underlying physics is not
understood, and so instead we will take a naive approach and set a prior on the
beam-widths from astrophysical data.

For the beam-widths, $\Wone$ and $\Wtwo$, we will use the same prior as defined
in Sec.~\ref{sec: noise-only fitting the model} for the noise-only model:
namely a uniform prior on $[0, 0.1\PATNF]$ covering 10\% of the spin-period.
Using such a prior introduces some ambiguity into the model comparison as the
result could be `tuned' by varying the fraction $f$ of the spin-period used to
set the uniform prior limits (here $f=0.1$). This issue is addressed in
Sec.~\ref{sec: effect of the choice of prior} where we find all sensible
choices of $f$ lead to the same overall conclusion.

The final parameter which
requires a prior distribution is the noise-component: as described in
Sec.~\ref{sec: noise-only} we apply a prior to $\sigma_{W_{10}}$ using a crude
estimate from the data; this is tabulated along with the other priors in
Table~\ref{tab: perera beam-width prior}.
\begin{table}
\centering
\caption{Prior distributions for the beam-width switching model. Parameters
for which the prior is taken from spin-down posteriors are labelled by $^{*}$.}
\label{tab: perera beam-width prior}
\input{Table_Beamwidth_Switching_0.1_prior.tex}
\end{table}

\begin{figure*}
\centering
\includegraphics[]{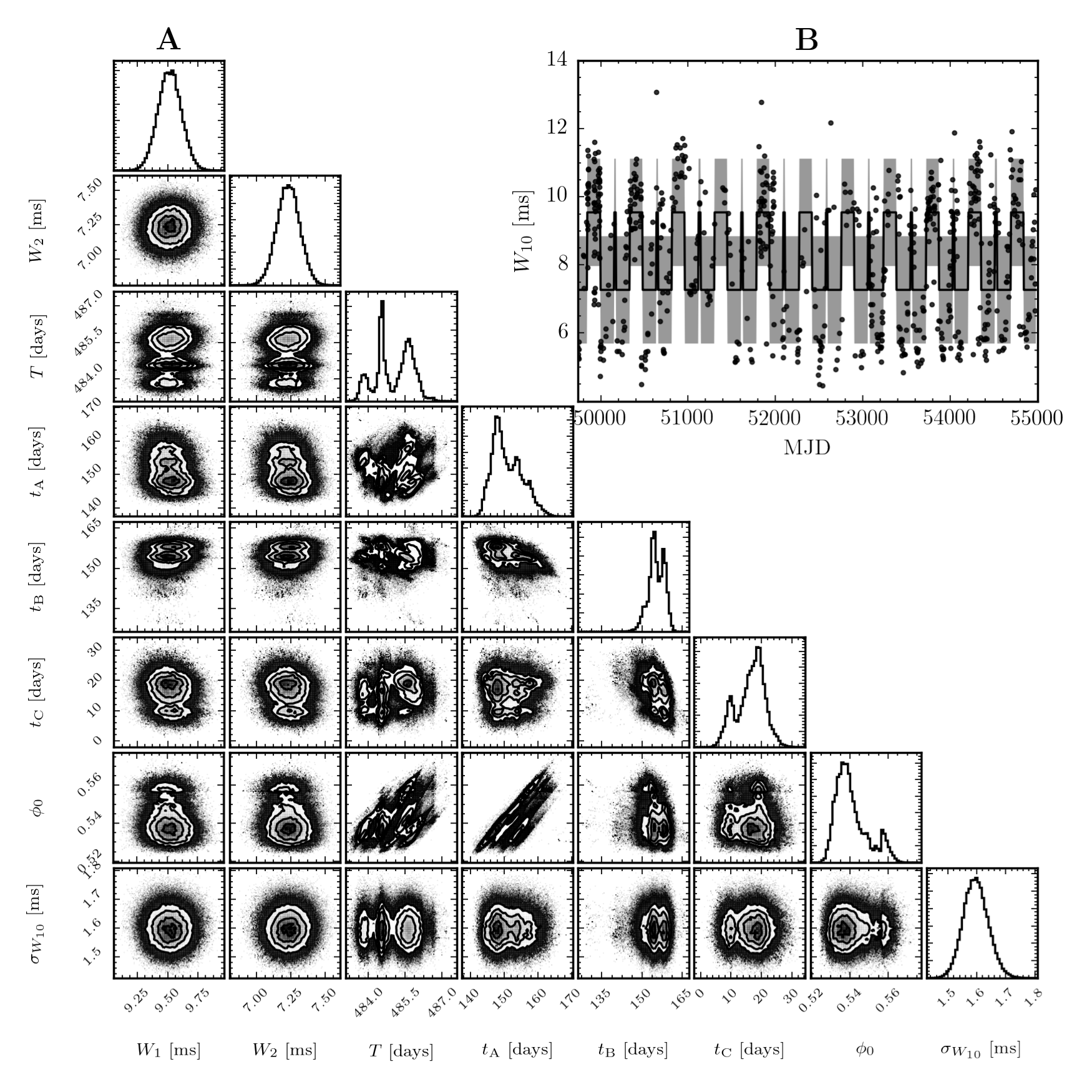}
\caption{\bigfigurecaptions{Switching}{beam-width}}
\label{fig: perera beam-width posterior}
\end{figure*}

We plot the posterior estimate
in Fig.~\ref{fig: perera beam-width posterior}A which demonstrates
non-Gaussianity and multimodal features in the segment times and the
phase. This indicates the existence of multiple solutions which could explain the
data. We note that the noise component $\sigma_{W_{10}}$ has a mode at 1.6 ms
which is less than the 2ms required in the noise-only model. This indicates
that some of the variability is being explained by the signal model.
In Table~\ref{tab: perera beam-width posterior} we summarise the posterior. We
find that the posterior modes satisfy $\Wone > \Wtwo$: larger
beam-widths are associated with the smaller absolute spin-down rates as found
by \mboxcitet{Lyne2010}.
\begin{table}
\centering
\caption{Posterior estimates for the beam-width switching model.}
\label{tab: perera beam-width posterior}
\input{Table_Beamwidth_Switching_0.1_posterior.tex}
\end{table}

Again we check the predictive power of our estimated posterior by plotting the
MPE alongside the data in Fig.~\ref{fig: perera beam-width posterior}B. The fit
to the data is not as good as the spin-down fit: by eye it is clear that most
data points lie away from the signal model requiring a greater (relative) level
of noise.

\subsection{Precession model}
\label{sec: The precession model}

We will now define the precession model and its predictions for the expected
signal in the spin-down and beam-width data.

Classical free precession refers to the rotation of a rigid non-spherical body
when there is a misalignment between its spin and rotation axes. For this work
we will consider a biaxial star, acted upon by an electromagnetic torque as
discussed in the next section. We will work with the angles defined in
\mboxcitet{Jones2001}: that is the star emits its EM radiation beam along the magnetic
dipole $\mathbf{m}$ which makes an angle $\chi$ with the symmetry axis of the
moment of inertia, and $\theta$ is the so-called wobble angle made between the
symmetry axis and the angular momentum vector. We will consider the
small-wobble angle regime where $\theta \ll 1$ since this is thought to be the
most physical solution for B1828-11.  Finally we define $P$ as the rotation
period, and $\dot P$ its time derivative, where the small variations due to
precession have been averaged over, and
\begin{align}
    \tauAge \equiv  \frac{P}{\dot P}.
\end{align}
as a characteristic spin-down age.

\subsubsection{Defining the spin-down rate model}

Observers infer the spin-down rate by measuring the arrival times of
pulsations.
For a freely precessing star the spin-down rate is periodically modulated on a
time-scale known as the \emph{free precession period}, which we will denote as
$\tauP$.  This result is referred to as the \emph{geometric modulation}
\citep{Jones2001} since it is a geometric effect.  Under the action of a torque
the geometric effect persists, but an additional electromagnetic effect enters
owing to torque variations \citep{Cordes1993}. The authors of
\mboxcitet{Jones2001} and \mboxcitet{Link2001} studied both effects in the
presence of a vacuum dipole torque \citep{Davis1970} and agreed that the
electromagnetic contributions dominate for B1828-11: we will therefore neglect
the geometric effect. The precession model for
B1828-11 has been developed by \citet{Akgun2006} where a non-vacuum
dipole torque was considered and additionally by \mboxcitet{Arzamasskiy2015}
where the effect of a plasma-filled magnetosphere was investigated. All of these are
potential areas of improvement, but in this work we will restrict our focus
to the simplest
specification capable of explaining the observations.  In particular, we will
use a generalisation of the vacuum dipole torque to allow for a braking index
$n\ne3$, but retain the angular dependence; this ansatz may be written as
\begin{equation}
\dot{\nu} = -k\nu^{n}\sin^{2}\Theta,
\end{equation}
where $k$ is a positive constant, and $\Theta$ is the
polar angle between the dipole and the angular momentum vector as
as calculated in Eqn.~(52) of \mboxcitet{Jones2001}.
Then we calculate \emph{secular} solutions in which $\Theta$ takes its fixed, time-averaged
value. We denote by $\theta$  the angle between the symmetry axis of the
moment of inertia and the angular momentum vector, and denote by $\chi$ the
angle between the symmetry axis and the magnetic dipole m.  We can then combine
the secular solution with an expansion of $\sin^2 \Theta$ in the small
$\theta$ limit, to give a spin-down rate
\begin{align}
\begin{split}
\dot{\nu}(t) = \frac{1}{\tauAge P}&\left(-1 + n\frac{1}{\tauAge} (t-\tref)
                \right. \\
& + \theta \left.\left[
2\cot\chi\sin\left(\psi(t)\right) - \frac{\theta}{2}\cos\left(2\psi(t)\right)
\right]\right),
\end{split}
\label{eqn: precession nudot degen}
\end{align}
where
\begin{equation}
\psi(t)= 2\pi\frac{t-\tref}{\tauP} + \psi_{0}.
\end{equation}
As in the switching model, $\tref= \textrm{MJD } 49621$.
The first two terms are the secular spin-down rate and its first derivative.
The term in the square brackets is the modulation and can be found from appropriate
manipulation of Eqn.~(58) and (73) in \mboxcitet{Jones2001}, or Eqn.~(20) in
\mboxcitet{Link2001}, aside from a factor of $\chi$ in the harmonic term which we
believe to be a misprint.

For $\chi < \pi/2$ the spin-down rate modulations are sinusoidal. When $\chi
\approx \pi/2$ (such that the star is nearly an orthogonal rotator), we will
see a strong harmonic at twice the precession frequency. It is precisely this
behaviour which is able to explain the doubly-peaked spin-down rate for
B1828-11.

\subsubsection{Parameter estimation for the spin-down}

In Table~\ref{tab: precession spin-down prior} we list the priors selected for
the spin-down precession model.
\begin{table}
\centering
\caption{Prior distributions for the spin-down precession model.}
\label{tab: precession spin-down prior}
\input{Table_Spindown_Precession_prior.tex}
\end{table}
For $\tauP$ and
$\sigma_{\dot{\nu}}$ we use a prior based on a crude estimate from the data.
For the spin-down age, braking-index, and pulse period we use a normal prior
taking the mean and standard deviation from B1828-11 measurements reported in the
ATNF catalogue: the values are listed in Table~\ref{tab: ATNF}. In the analysis,
we present the posteriors of the difference to the means of these values for
convenience.
\begin{table}
\centering
\caption{Measured and inferred values of the precession spin-down model parameters
from the ATNF pulsar catalogue \citep{ATNF}, these are given at epoch MJD~49621}
\label{tab: ATNF}
\begin{tabular}{ll} \hline
$P^{\mathrm{ATNF}}$ & 0.405043321630 $\pm 1.2\times10^{-11}$~s \\
$n^{\mathrm{ATNF}}$ & 16.08 $\pm$ 0.17 \\
$\tauAge^{\mathrm{ATNF}}$ & 213827.91 $\pm 0.32$~ yrs\\
\end{tabular}
\end{table}
Finally for $\psi_0$ we give the full domain of
possible values. Since our derivation of the signal models in Sec.~\ref{sec:
The precession model} assumed the small wobble-angle regime $\theta \ll \chi$
and $\chi \sim \pi/2$, we similarly restrict their uniform priors to the
relevant range.

\begin{figure*}
\centering
\includegraphics[]{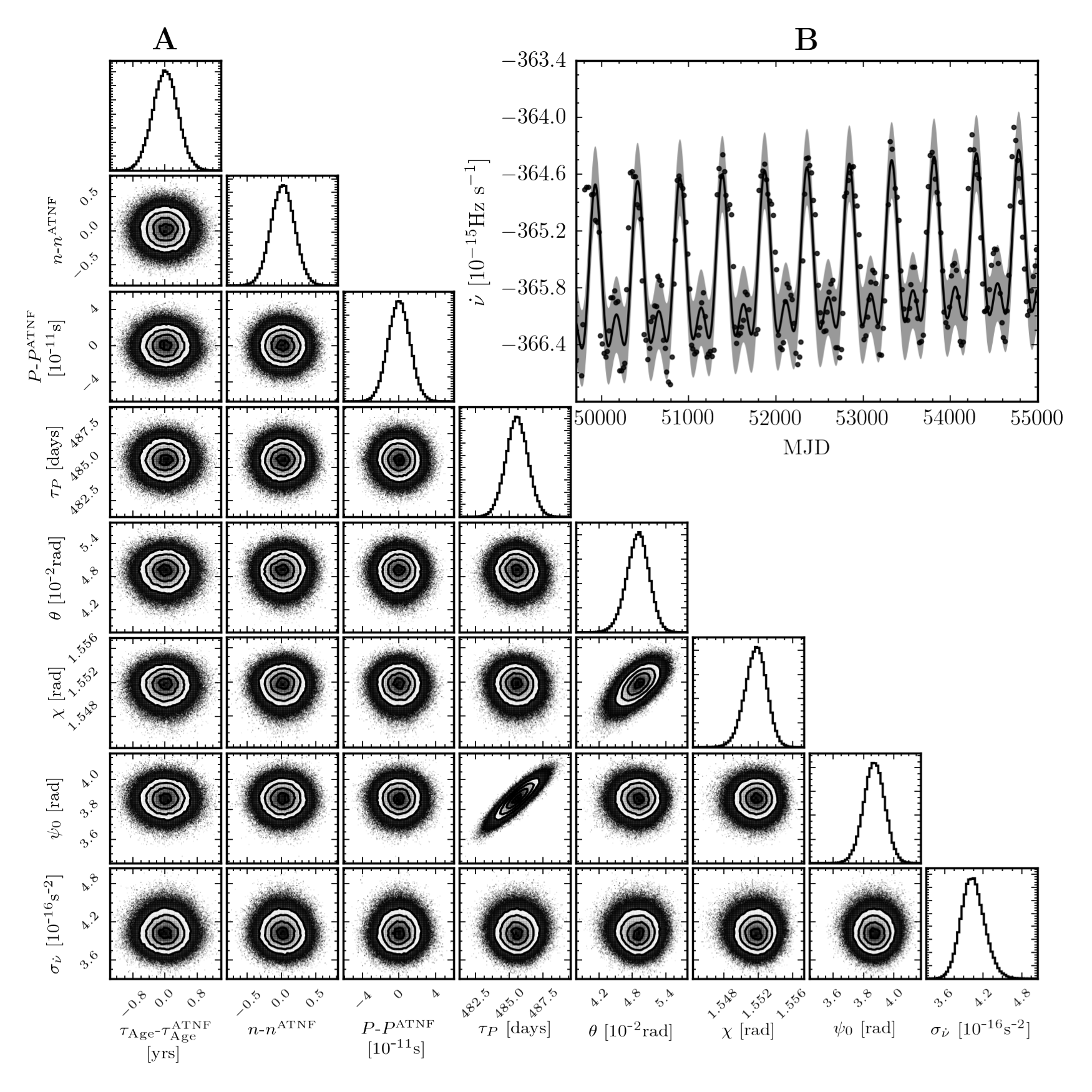}
\caption{$\textbf{A}$: The estimated marginal posterior probability distribution for the
precession spin-down model parameters. For the secular spin-down
quantities, we show the difference with respect to the values as listed in
table~\ref{tab: ATNF}.
$\textbf{B}$: Checking the fit of the model using the
maximum posterior values to the data; see Fig.~\ref{fig: noise-only beam-width
posterior fit} for a complete description.}
\label{fig: precession spin-down posterior}
\end{figure*}
Running the MCMC simulations we plot the resulting
posterior in Fig.~\ref{fig: precession spin-down posterior} and provide a
summary in Table~\ref{tab: precession spin-down posterior}. For all parameters
the posterior distribution is Gaussian: in the case of parameters which we gave
a uniform prior, this indicates that the spin-down data is informative. For the
parameters using an informative prior (from the ATNF catalogue), the posterior
and prior are similar, this indicates the data agrees with the prior, but does
not significantly improve our estimates.
\begin{table}
\centering
\caption{Posterior estimates for the spin-down precession model. For the secular spin-down
quantities, we report the posterior difference with respect to the values as listed in
table~\ref{tab: ATNF}.}
\label{tab: precession spin-down posterior}
\input{Table_Spindown_Precession_posterior.tex}
\end{table}

In Fig.~\ref{fig: precession spin-down posterior}B we check the fit of the
posterior for the spin-down data. The spin-down model fits to the data points
well with only a small amount of noise required to explain the data.

The posterior distributions conditioned on the spin-down data (as summarised in
Table~\ref{tab: precession spin-down posterior}) can be compared with the
values reported in Table~2 of \mboxcitet{Link2001}. When comparing it should be
noted that we are considering a longer stretch of data which includes most, but
not all of the period studied by \mboxcitet{Link2001}. For the two angles $\chi$
and $\theta$, the fractional difference is 0.001 and 0.14 respectively while
the precession periods differ by a fractional amount 0.05. Clearly the solution found here
is similar to that found by \mboxcitet{Link2001}.

\subsubsection{Defining the beam-width model: Gaussian intensity}

\begin{figure}
\centering \includegraphics[width=0.2\textwidth]{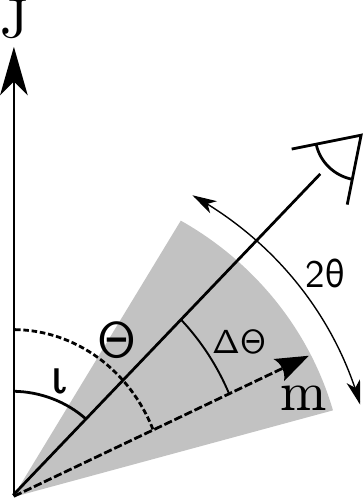}
\caption{Illustration of the angles as the beam-axis $\mathbf{m}$ cuts the
         plane containing the observer and the angular momentum~$\mathbf{J}$.
         The grey shaded region indicates the extent to which $\mathbf{m}$
         varies over a precessional cycle.}
\label{fig: J observer plane}
\end{figure}

Modulation of the observed beam due to precession is a purely geometric effect.
Fixing the beam-axis to coincide with the magnetic dipole $\mathbf{m}$ and
following \mboxcitet{Jones2001}, we define $\Theta$ and $\Phi$ as the polar and
azimuthal angles of $\mathbf{m}$ with respect to a fixed Cartesian coordinate
system with $z$ along the angular momentum vector $\textbf{J}$. The observer is
fixed in the Cartesian coordinate system with a polar angle $\iota$ to $\textbf{J}$, and
azimuth $\PhiO$. The slow precessional motion of the spin-vector causes modulation
in the angle $\Theta$:
\begin{align}
\Theta(t) = \cos^{-1}\left(\sin\theta\sin\chi\sin\psi(t) + \cos\theta\cos\chi\right),
\label{eqn: Theta}
\end{align}
which, in the $\theta \ll 1$ limit is approximately
\begin{align}
\Theta(t) \approx \chi - \theta \sin\psi(t).
\end{align}
Taking the plane containing the angular momentum vector and the observer, in
Fig.~\ref{fig: J observer plane} we demonstrate the range of motion of
$\mathbf{m}$ over a precessional cycle by a gray shaded region.  The region has
a mean polar value of $\chi$ and a range of $2\theta$.

\begin{figure}
\centering
\includegraphics{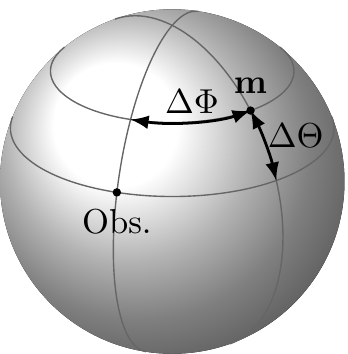}
\caption{The angular position of the observer and the magnetic dipole
         $\mathbf{m}$ on the unit sphere centered on the star; $\Delta\Theta$
         and $\Delta\Phi$ are then the polar and azimuthal angles between
         them.}
\label{fig: delta theta delta phi}
\end{figure}

For an observer fixed at an angle $\iota$ to $\textbf{J}$ we define an impact parameter:
\begin{align}
\label{eqn:Delta_Theta_definition}
\Delta\Theta(t) = \Theta(t) - \iota ,
\end{align}
which will vary in time with the precession period $\tauP$.  This impact
parameter determines how the observer's line-of-sight cuts the emission beam;
if $\Theta(t)$ varies due to changes in $\Delta\Theta(t)$, then the observer
will measure the beam to vary on the slow precession time-scale. To help visualise the setup, in
Fig.~\ref{fig: delta theta delta phi} we plot the unit sphere with points
corresponding to the beam-axis $\mathbf{m}$ and the observer. For each of these
we have added lines of latitude and longitude. Then we see that that
$\Delta\Theta$ is the difference between the lines of latitude and we can also
define $\Delta\Phi(t) = \Phi(t) - \Phi_{\textrm{obs}}$ as the difference in the
lines of longitude.

The analysis by \mboxcitet{Stairs2000} characterised the beam by a shape parameter.
\mboxcitet{Link2001} used an expansion in $\Delta\Theta$ to model the beam-width
and hence the shape-parameter. This allowed them to use their fit to the
timing-data to infer the beam-geometry which they found to be hour-glass
shaped; see their Figure 5 for a schematic illustration.   The authors of
\mboxcitet{Lyne2010} did not use a shape-parameter as it requires time-averaging
over a longer base-line, something they wish to avoid in order to be able to
observe the switching. Instead, they considered the beam-width at 10\% of the
maximum, $W_{10}$, which is measured on a shorter time-baseline ($\sim1$~hr).
If we want to use the beam-width to make a model comparison, we will require a
model for $W_{10}$ that is not informed by the data.

The integrated pulse profile of B1828-11 (Fig. 4 of \mboxcitet{Lyne2010})
shows a single peak, often described as  core emission \mbox{\citep{Lyne1988}}.  Since
we do not have a detailed model of the emission mechanism, we will now consider
the most rudimentary and natural beam geometry which fits this: a circularly
symmetric (about the beam-axis) intensity which falls-of with a Gaussian
function.  Specifically, let us define $\Delta d$ as the central angle between
the observer's line-of-sight and the beam (this is the spherical distance
between the two points marked in Fig.~\ref{fig: delta theta delta phi}), then
the intensity is
\begin{equation}
\In(t) = \In_{0}\exp\left(-\frac{\Delta d(t)^{2}}{2\rho^{2}}\right).
\label{eqn: beam intensity}
\end{equation}
Here $\In_0$ is the intensity when observed directly along the dipole and
$\rho$ measures the angular width of the beam.
From the spherical law of cosines, for an observer located at $(\PhiO, \iota)$,
we have
\begin{align}
\Delta d(t) &= \cos^{-1}\left[\cos\Theta(t)\cos\iota +
                              \sin\Theta(t)\sin\iota\cos|\Delta\Phi(t)|\right].
\label{eqn: angular sep}
\end{align}
The observer will see a maximum pulse intensity at $\Delta\Phi = 0$, given by
\begin{align}
\In_\textrm{max} = \In_{0} \exp\left(
    -\frac{\left(\Theta(t) - \iota\right)^{2}}
          {2\rho^{2}}\right).
\label{eqn: imax}
\end{align}

Now let us recognise that $\Theta$ varies on the slow precession time-scale,
while $\Phi$ varies on the rapid spin time-scale: a single pulse consists of
$\Phi$ varying between $\PhiO - \pi$ and $\PhiO + \pi$. So over a single pulse,
we can treat $\Theta$ as a constant. The pulse width $W_{10}$, as measured by
observers, is the duration for which the pulse intensity is greater than 10\%
of the peak observed intensity. For a single pulse, we can define this duration
as the period for which the inequality
\begin{align}
\In > \In_{\textrm{max}} \frac{1}{10},
\label{eqn: beam-width ineq}
\end{align}
is satisfied.  Substituting equations~\eqref{eqn: beam intensity}
and~\eqref{eqn: imax} into \eqref{eqn: beam-width ineq} and rearranging we find
that
\begin{equation}
\cos(|\Delta\Phi(t)|) > \frac{\cos\Psi(t)-\cos\Theta(t)\cos\iota}
    {\sin\Theta(t)\sin\iota},
\label{eqn: cos delta phi}
\end{equation}
where
\begin{align}
\Psi(t) = \sqrt{
    \left(\Theta(t) - \iota\right)^{2} +
     2\rho^{2} \ln\left(10\right)}.
\label{eqn: Psi}
\end{align}
Since we treat $\Theta$ as a constant over a single pulsation, we can also treat
the whole right-hand-side of the inequality as a constant during each pulse.

Now consider a single rotation with the magnetic dipole starting and ending in
the antipodal point to the observer's position such that $\Delta\Phi(t)$ increases
between $-\pi$ and $\pi$ during this rotation. Then inequality~\eqref{eqn: cos delta phi} measures the fraction of the pulse corresponding to
the beam-width measurement. In terms of the rotation, we define $\delta\Phi$ as
the angular width for which the inequality is satisfied and calculate it to be
\begin{align}
\delta\Phi(t) = 2\cos^{-1}\left(
\frac{\cos\Psi(t)-\cos\Theta\cos\iota}
    {\sin\Theta\sin\iota}
\right).
\end{align}
Then the beam-width is
\begin{align}
W_{10}(t) = P \frac{\delta\Phi(t)}{2\pi},
\end{align}
from which we arrive at
\begin{align}
W_{10}(t) = \frac{P}{\pi} \cos^{-1}\left(
\frac{\cos \Psi(t) - \cos\Theta(t)\cos\iota}{\sin\Theta(t)\sin\iota}
\right).
\label{eqn: w10 beam-width gaussian}
\end{align}

In order for the observer to measure the width at 10\% of the maximum, the beam
intensity must of course drop below this value before increasing again. In
reality we typically observe pulse durations lasting for small fractions of the
period, especially when they are close to orthogonal rotators \citep{Lyne1988}.

To set a prior on $\rho$ we consider a special case in which the polar angle of
the beam and the observer are at the equator ($\Theta=\iota=\pi/2$). From our
spin-down analysis, we know the first of these conditions is true for B1828-11
since $\chi$ is close to $\pi/2$. The second condition is based on the assumption
that the observer would not see a tightly pulsed beam if they are not close to the
polar angle of the beam. In this special instance, inserting
Eqn.~\eqref{eqn: Psi} into Eqn.~\eqref{eqn: w10 beam-width gaussian}, the beam-width is
\begin{align}
W_{10} \bigg|_{\Theta=\iota=\pi/2} = \frac{P}{\pi}\sqrt{2\ln10} \rho.
\end{align}
To set a prior on $\rho$, we can equate this with the beam-widths used in the
switching model, for which we set a uniform prior from $0$ to $0.1\PATNF$.
To make an even-handed comparison we will therefore set a uniform prior on
$\rho$ from 0 to
\begin{align}
\frac{\pi}{10\sqrt{2\ln10}} \approx 0.15,
\label{eqn: rho prior}
\end{align}
so that, for this special case, the prior range of $\rho$ corresponds exactly
to the prior range of the beam-widths in the switching model.
This prior range will change, but not by orders of magnitude when
considering a system close to, but not exactly at, this special case. Therefore
this prior assures that the model comparison does not introduce any significant
bias into the model comparison.

\subsubsection{Parameter estimation for the Gaussian beam-width model}

We are in a position to fit the Gaussian beam model to the observed $W_{10}$
values. In Table~\ref{tab: gaussian beam-width prior} we list the priors taken
from the spin-down fit along with three additional priors. For $\iota$ we
choose a uniform prior in $\cos\iota$ on $[-1, 1]$, this corresponds to
allowing $\iota$ to range from $[0, \pi]$ (the observer could be in either
hemisphere); for $\rho$ we apply the prior from Eqn.~\eqref{eqn: rho
prior}; and for $\sigma_{W_{10}}$ we use a crude estimate based on the
data (again we use the same prior for all three models).
\begin{table}
\centering
\caption{Prior distributions for the beam-width Gaussian precession model. Parameters
for which the prior is taken from spin-down posteriors are labelled by $^{*}$.}
\label{tab: gaussian beam-width prior}
\input{Table_Beamwidth_Gaussian_prior.tex}
\end{table}

Fitting Eqn.~\eqref{eqn: w10 beam-width gaussian} to the data we discover that
the Gaussian beam model is a poor fit to the data. In Fig.~\ref{fig: gaussian
beam-width posterior}B the MPE shows that while the model is able to fit the
averaged beam-width, it cannot simultaneously fit the amplitude of periodic
modulations.
\begin{figure*}
\centering
\includegraphics[]{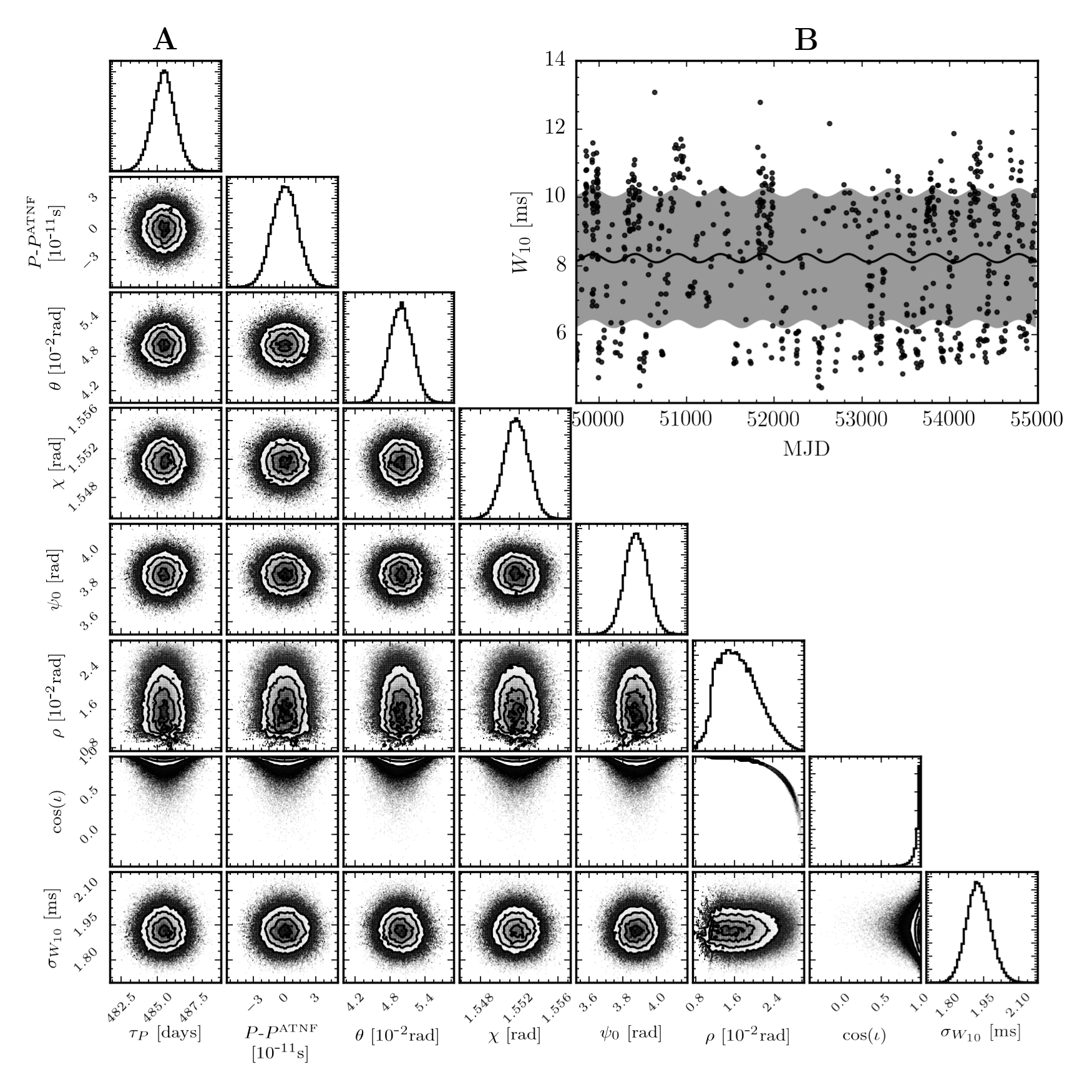}
\caption{\bigfigurecaptions{Gaussian}{spin-down}}
\label{fig: gaussian beam-width posterior}
\end{figure*}

The posterior distribution (as seen in Fig.~\ref{fig: gaussian beam-width
posterior}A) is Gaussian for all of the parameters except $\cos\iota$ for which
it concentrates the probability at $\iota\approx0$: the observer looks almost
down the angular momentum vector. Since $\chi\approx\pi/2$ and $\theta \ll 1$,
for each pulsation the beam must therefore sweep out a cone with such a large opening angle
it is close to a plane orthogonal to the rotation vector. Meanwhile the
rotation vector is nearly parallel to the angular momentum, since $\theta \ll
1$. As a result, the beam remains approximately orthogonal to the observer for
the entirety of each pulsation.  We find this result difficult to believe on
the grounds that the observer would \emph{not} see tightly coliminated pulsed
emission.  For this reason, we conclude that the Gaussian beam-intensity fails
to fit the data because the best-fit is unphysical.
In retrospect, this result is not surprising since a Gaussian beam intensity is
known to have a beam-width (as measured by $W_{10}$) which is \emph{independent} of
the impact angle as discussed by \mboxcitet{Akgun2006}. This is a direct result of
measuring the beam-width with respect to the observed maximum and the self-similar
nature of the Gaussian intensity under changes in the impact parameter.

\subsubsection{Refining the beam-width model: modified-Gaussian intensity}
\label{sec: refining the beam-width model}

As we have demonstrated that the Gaussian beam is unable to explain both the
observed variations and average beam-width we must now consider how it could be
varied in a natural way which does explain the data. One suggestion from
\mboxcitet{Akgun2006} is to impose a sharper cut-off, or introduce a conal
component in addition to the Gaussian core emission.  We will follow a slightly
different path below, one which represents a less drastic modification of the
beam profile.


The beam intensity described by Eqn.~\eqref{eqn: beam intensity} is circularly
symmetric about the beam axis as viewed on the surface of the sphere. In the
context of the hollow-beam model \citep{Radhakrishnan1969}, \mboxcitet{Narayan1983}
found that pulsar beams can be elongated with the ratio of major to minor axis
being $\sim3$ for typical pulsars. B1828-11 does not fit into the hollow-beam
model (having only a single core component), but nevertheless if the conal
emission can be non-circular a generalisation of our core intensity would be
to allow for an elliptical beam.

To consider non-symmetric geometries, let us take the planar limit of
Eqn.~\eqref{eqn: angular sep} by
applying small angle approximations in $\Delta d, \Delta\Theta$ and
$\Delta\Phi$:
\begin{align}
\Delta d(t)^{2} = \Delta \Theta(t)^{2} + \sin\Theta(t) \sin\iota \Delta\Phi(t)^{2}.
\label{eqn: angular sep expansion}
\end{align}
This corresponds to setting the observer close to $\mathbf{m}$ in
Fig.~\ref{fig: delta theta delta phi}.

Obviously $\Delta\Phi$ ranges over $[0,2\pi]$ in each rotation, but when
$\Delta\Phi$ is not small, the intensity vanishes rapidly due to the Gaussian
beam shape Eqn.~\eqref{eqn: imax}.  Therefore, Eqn.~\eqref{eqn: angular sep
expansion} is a good approximation for the separation when the beam is pointing
near to the observer, while away from this it is a poor approximation, but the
intensity is negligible and so the differences are inconsequential.

We can now allow for an elliptical beam geometry by postulating the beam
intensity to be
\begin{align}
\In(t) = \In_{0} \exp\left(
-\frac{-\Delta\Theta(t)^{2}}{2\rho_1^{2}}
-\frac{(\sin\Theta(t)\sin\iota\Delta\Phi(t))^{2}}{2\rho_2^{2}}
\right).
\label{eqn: 2D intensity}
\end{align}
Then to calculate the beam-width, we first find the maximum:
\begin{align}
\In_\textrm{max} = \In_{0} \exp\left(
-\frac{-\Delta\Theta(t)^{2}}{2\rho_1^{2}}
\right).
\end{align}
Solving for the beam-width we find
\begin{align}
W_{10}(t) = \frac{P}{\pi}\frac{\sqrt{2\ln10}\rho_{2}}{\sin\Theta(t)\sin\iota},
\label{eqn: W10 elliptical}
\end{align}
which is independent of $\rho_1$, the latitudinal standard-deviation. The extra
degree of freedom introduced in Eqn.~\eqref{eqn: 2D intensity} is irrelevant to
the beam-width measure because $W_{10}$ is defined by the ratio of the
intensity to that at the observed peak $\In_{\textrm{max}}$.

This loss of a degree of freedom means that Eqn.~\eqref{eqn: W10 elliptical} is
an equivalent to an expansion of Eqn.~\eqref{eqn: w10 beam-width gaussian} in
the planar limit (i.e. the non-circular nature introduced by Eqn.~\eqref{eqn: 2D intensity}
does not manifest in the prediction for $W_{10}$)
and so will suffer the same problems if fitted to the data. To further
generalise our intensity model we will therefore modify the beam-geometry by
allowing a varying degree of  non-circularity. This is done by expanding the
longitudinal standard deviation as
\begin{align}
\rho_{2}(t) = \rho_2^{0} + \rho_2'' \Delta\Theta(t)^{2}.
\label{eqn: rho2}
\end{align}
Note that we have neglected to include a linear term here, forcing the geometry
to be longitudinally symmetric about the beam-axis. Preliminary studies began
by fitting a linear term only (this giving a modulation at the frequency
$1/\tauP$), but it was found that including a second-order term (which provides
modulation at both $1/\tauP$ and $2/\tauP$) gave a better fit. Including both
terms, we found that the data was unable to provide inference on both $\rho_2'$
and $\rho_2''$ due to degeneracy. In light of this, we drop the first
term, but keep the second, which we feel is the simplest model which is able to
fit the data.

Solving for the beam-width (i.e. with Eqn.~\eqref{eqn: rho2} substituted into
Eqn.~\eqref{eqn: 2D intensity}) we obtain a signal model
\begin{align}
W_{10}(t) = \frac{P}{\pi}\sqrt{\frac{2\ln10}{\sin\Theta(t)\sin\iota}}
 \left(\rho_2^{0} + \rho_2'' \Delta\Theta(t)^{2} \right),
\label{eqn: signal model jones}
\end{align}
which we will refer to as the \emph{modified-Gaussian precession beam-width} model.

\subsubsection{Parameter estimation for the modified-Gaussian precession
               beam-width}

For equation \eqref{eqn: signal model jones}, we give the relevant prior distributions
in Table~\ref{tab: modified-gaussian beam-width prior}. As in the previous
Gaussian model, we let $\iota$ range over $[0, \pi]$; for $\rho_2^{0}$, we
apply the prior on intensity widths as given by Eqn.~\eqref{eqn: rho
prior}; and for $\rho_2''$ we will use a normal prior with zero mean
favouring a Gaussian intensity. The \emph{standard-deviation} of this prior can
have a measurable impact on the inference: if it is too small then the degree
of freedom introduced by Eqn.~\ref{eqn: rho2} is effectively removed. Instead
we want to make it significantly larger than the (apriori unknown) posterior
value of $\rho_2''$: this generates a so-called non-informative prior. To set
the prior standard-deviation then, we need to provide a rough scale for what
value $\rho_{2}''$ should have. To do this we will define our prior expectation
such that
\begin{align}
\rho_2(\Delta\Theta = \rho_2^0) \sim 2\rho_2^0,
\label{eqn: rho2 prior}
\end{align}
which is to say we expect $\rho_2$ to increase by no more than a factor of
order unity over angular distances of the beam-width comparable to $\rho_2^0$
(the beam-width when the observer cuts directly through the beam-axis). This
amounts to assuming that the beam does not depart very far from circularity.
Plugging this into Eqn.~\eqref{eqn: rho2}, we get
\begin{align}
\rho_2'' \sim \frac{1}{\rho_2^0}.
\label{eqn: rho2dd prior}
\end{align}
From this, we use the upper limit from the uniform prior on $\rho_2^0$ (as calculated in
Eqn.~\eqref{eqn: rho prior}), to set the
standard-deviation for $\rho_2''$ at $1/0.15\approx7$.
We also tested different choices of $\rho_2''$ and found that the
posteriors and odds-ratios where robust to the choice, provided the
standard-deviation did not exclude the posterior value reported in
table~\ref{tab: modified-gaussian beam-width posterior}.
\begin{table}
\centering
\caption{Prior distributions for the beam-width modified-Gaussian precession
model. Parameters for which the prior is taken from spin-down posteriors are
labelled by $^{*}$.}
\label{tab: modified-gaussian beam-width prior}
\input{Table_Beamwidth_ModifiedGaussian_0.1_prior.tex}
\end{table}

\begin{figure*}
\centering
\includegraphics[]{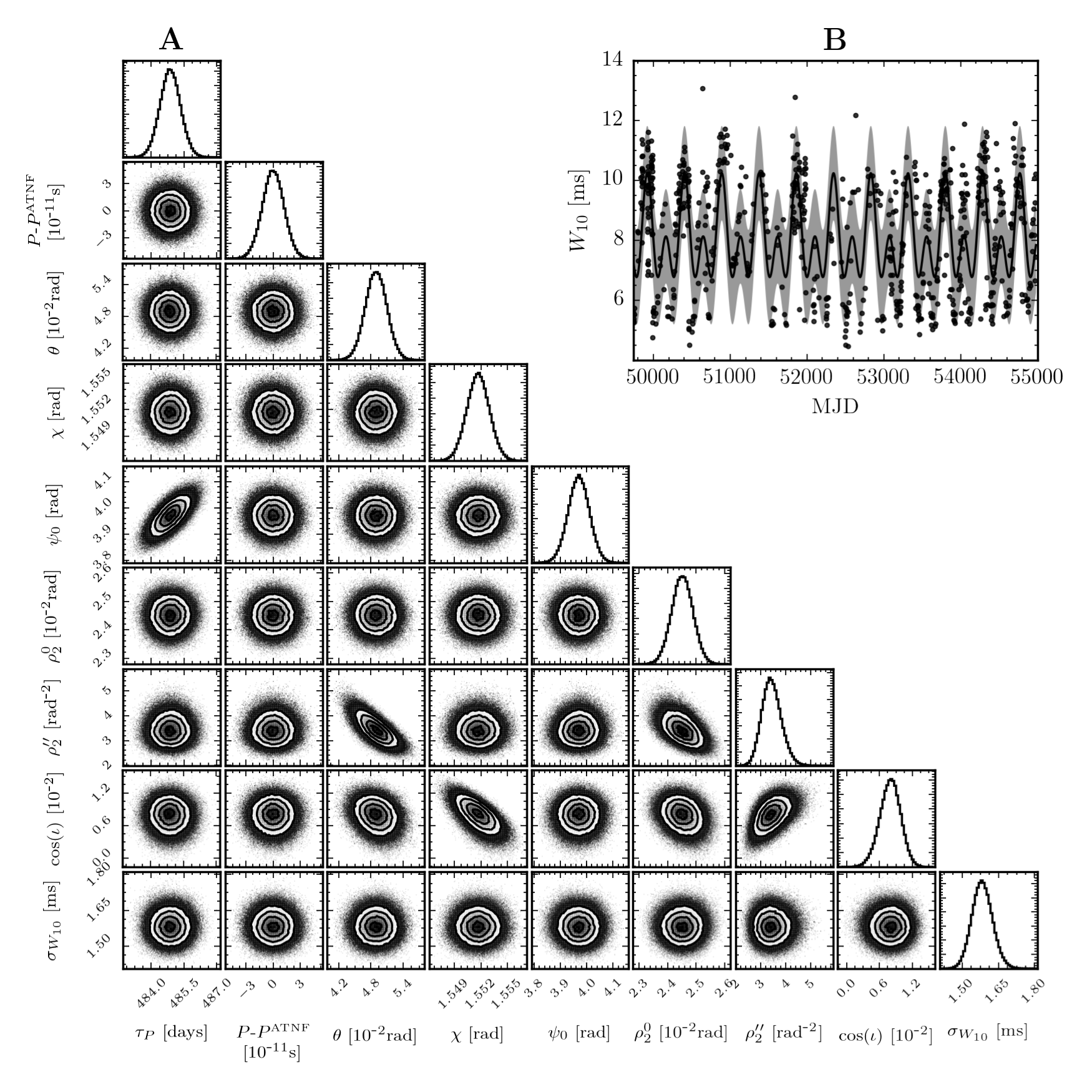}
\caption{\bigfigurecaptions{modified-Gaussian precession}{beam-width}}
\label{fig: modified-gaussian beam-width posterior}
\end{figure*}

The MCMC simulations converge quickly to a Gaussian distribution as shown in
Fig.~\ref{fig: modified-gaussian beam-width posterior}A and the posterior is
summarised in Table.~\ref{tab: modified-gaussian beam-width posterior}. The model parameters
common to the spin-down model do not vary significantly from the spin-down
posterior: this indicates the two models are consistent. We find that $\iota$
is close to $\pi/2$ as expected, $\rho_2$ is sufficiently small indicating a
narrow pulse beam, but $\rho_2''$ has departed from its prior mean of zero.
This confirms that our generalisation of the Gaussian intensity,
Eqn.~\eqref{eqn: rho2}, is important in fitting the data.
\begin{table}
\centering
\caption{Posterior estimates for the beam-width modified-Gaussian precession model.}
\label{tab: modified-gaussian beam-width posterior}
\input{Table_Beamwidth_ModifiedGaussian_0.1_posterior.tex}
\end{table}

In Fig.~\ref{fig: modified-gaussian beam-width posterior}B we perform the
posterior predictive check plotting the MPE alongside the data.  This
demonstrates that the best-fit puts $\chi$ within $\theta$ of $\iota$ such that
during the precessional cycle the beam-axis passes twice through observer's
location. This corresponds to the gray region in Fig.~\ref{fig: J observer
plane} intersecting the observer's line-of-sight. When this happens, the
modulation of the beam-width picks up a second harmonic at twice the precession
frequency. The minima in Fig.~\ref{fig: modified-gaussian beam-width
posterior}B corresponds to the point in the precessional phase when the
beam-axis point directly down the observer's line-of-sight.

\subsubsection{Recreating the beam-geometry}
\label{sec: recreating the beam-geometry}

Since we have defined a beam-intensity in Eqn.~\eqref{eqn: 2D intensity} we can
recreate the beam-geometry and pulse-shape from our MPE values.  The data we
have does not provide information about the latitudinal beam-shape parameter
$\rho_1$; therefore we consider that there are a family of beam-geometries
parametrized by $\rho_1 = \lambda \rho_2^0$ where $\lambda$ is an arbitrary
scale parameter and $\rho_2^0$ is the MPE value.

\begin{figure}
\centering
\includegraphics[width=0.48\textwidth]{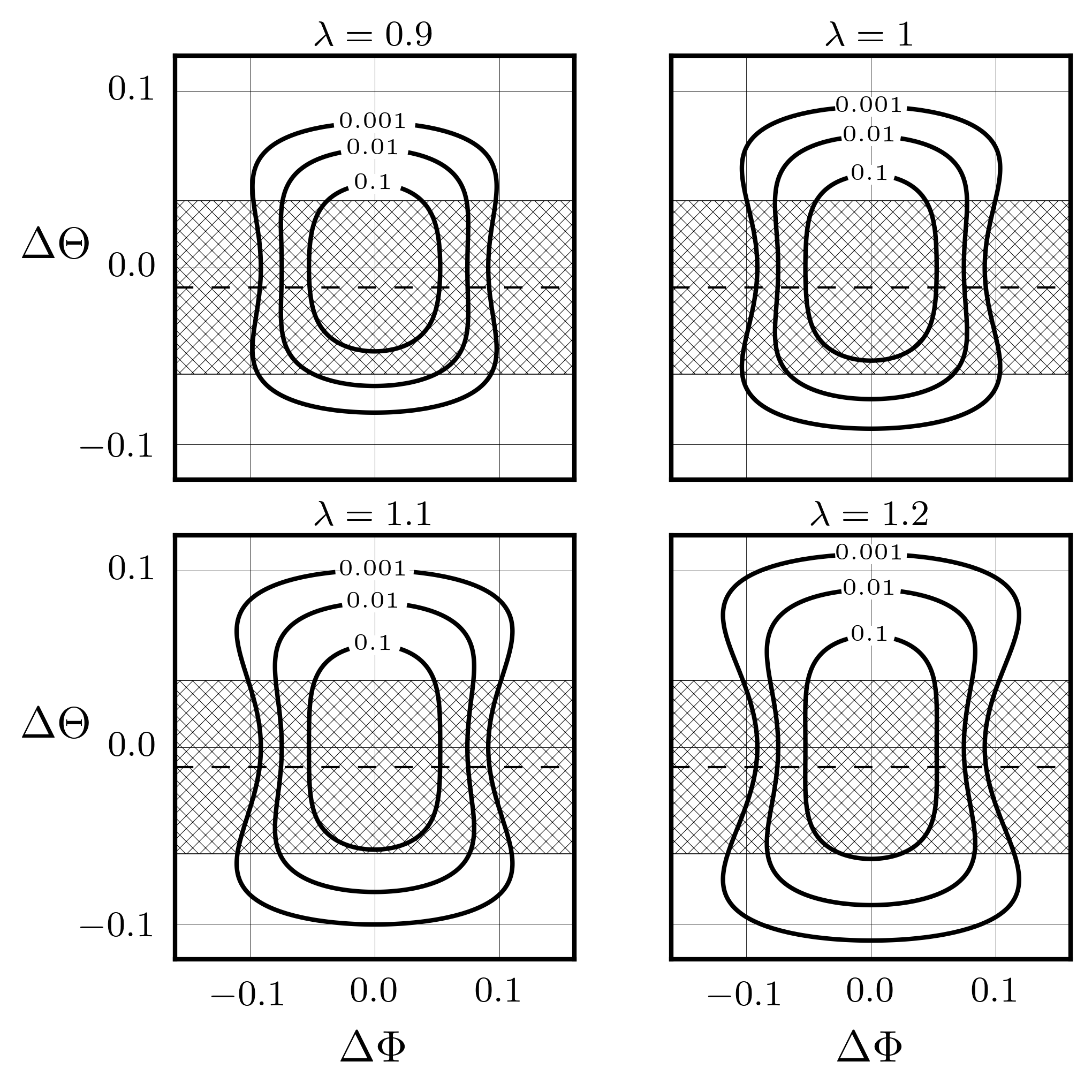}
\caption{Recreating the beam geometry from the MPE of the modified-Gaussian
precession beam-width model parameters for four different values of $\lambda$.
Thick black lines indicate contour lines of the intensity function at fractions
of the maximum intensity. The hatched area indicates the region of horizontal
cuts (pulses) sampled by the observer: this has a mean, $\chi-\iota$, close to
zero (marked by a dashed line) and varies by $\pm\theta$ about this mean.}
\label{fig: modified-gaussian beam-width recreate geometry}
\end{figure}

In figure \ref{fig: modified-gaussian beam-width recreate geometry} we pick
four illustrative values for $\lambda$ and plot the resulting beam-geometry as
contour lines at fixed fractions of the maximum beam intensity (which occurs at
the origin).  This demonstrates that the beam-geometry has an hour-glass shape
in agreement with \mboxcitet{Link2001}, although this becomes weaker with smaller
values for $\lambda$.

In Fig.~\ref{fig: modified-gaussian beam-width recreate geometry}, a pulse
corresponds to a horizontal cut through the intensity at fixed $\Delta\Theta$.
Our posterior distribution, Fig.~\ref{fig: modified-gaussian beam-width
posterior}A, also provides information on how the observations cut through this
beam-geometry.  Under the precession hypothesis, the observer has a
time-averaged $\Delta\Theta$ of $\chi - \iota$: this has been plotted as a
horizontal dashed line in Fig.~\ref{fig: modified-gaussian beam-width recreate
geometry}. Precession modulates $\Delta\Theta$ about this average value by $\pm
\theta$; the observer's line-of-sight through the beam therefore varies by
$2\theta\approx0.1$~rad over a precessional cycle. We have plotted a gray
shaded region in Fig.~\ref{fig: modified-gaussian beam-width recreate geometry}
to show the extent, $\chi-\iota \pm \theta$, over which $\Delta\Theta$ varies
during a precessional cycle.

We stress here that the contour lines cannot be used directly to measure the
beam-width $W_{10}$. This is because $W_{10}$ is defined as the width at 10\%
of the peak intensity for that observed pulse and not the maximum intensity of
the beam.  The peak intensity for an observed pulse (a horizontal slice) is the
intensity at $\Delta\Phi=0$ and it is with respect to this, which $W_{10}$ is
measured.

By construction, the four beam geometries in Fig.~\ref{fig: modified-gaussian
beam-width recreate geometry} all produce the same $W_{10}$ behaviour as
observed in Fig.~\ref{fig: modified-gaussian beam-width posterior}B. The reason
for this is that we have lost information on the total intensity by using
$W_{10}$; other measurements of the beam-width could potentially yield more information
and better constrain the beam geometry.

\begin{figure}
\centering
\includegraphics[width=0.48\textwidth]{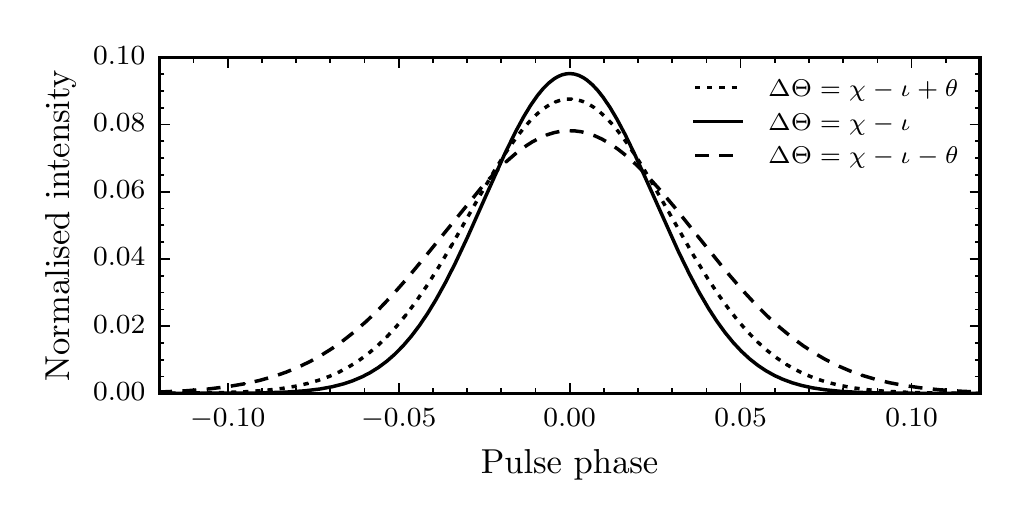}
\caption{Recreating the pulse profiles for three particular slices through the
beam using a fixed value of $\lambda=1$.}
\label{fig: modified-gaussian beam-width recreate pulse}
\end{figure}

Fixing $\lambda=1$ we can also consider the variations in the pulse profile. In
Fig.~\ref{fig: modified-gaussian beam-width recreate pulse} we plot the
normalised intensity for three values of $\Delta\Theta$ corresponding to the
mean, and edges of the gray region in Fig.~\ref{fig: modified-gaussian
beam-width recreate geometry}.
This figure shows that the narrow beam-widths occur when $\Delta\Theta$ is
small, which, since $\chi$ is close to $\pi/2$ coincide with the larger
(absolute) spin-down rates. This agrees with the findings of \mboxcitet{Lyne2010}
and this figure can be directly compared with panel C in Fig.~3 of that work.

\section{Estimating the odds-ratio}
\label{sec: estimating the odds-ratio}

\subsection{Thermodynamic integration}

Having checked that our MCMC simulations are a reasonable approximation to the
posterior distribution we now calculate the marginal likelihood for each model
and then their odds-ratio. To calculate the marginal likelihood we will use
thermodynamic integration. This requires running $N$ parallel MCMC simulations
and raising the likelihood to a power $1/T$ where $T$ is the `temperature' of
the chain. This method was originally proposed by \mboxcitet{Swendsen1986} to
improve the efficiency of MCMC simulations for multimodal distributions.
In this work we use this method not to help with the efficiency of the
simulations \footnote{All the posteriors are either unimodal or multimodal with
little separation between the modes}, but instead so that we can apply the
method prescribed by \mboxcitet{Goggans2004} to estimate the evidence as follows.

First we define the inverse temperature $\beta = 1/T$, then let the marginal
likelihood as a function of $\beta$ be
\begin{equation}
Z(\beta) = \int P(\data| \params)^{\beta}P(\params) d\params.
\end{equation}
When $\beta=1$, this gives exactly the marginal likelihood first defined in
Eqn.~\eqref{eqn: marginal likelihood}.
After some manipulation we see that
\begin{equation}
\frac{1}{Z} \frac{\partial Z}{\partial \beta} =
\frac{
\int \ln(P(\data| \params) P(\data| \params)^{\beta}P(\params) d\params}
{\int P(\data| \params)^{\beta}P(\params) d\params}.
\end{equation}
From this, we note that the right-hand-side is an average of the log-likelihood
at $\beta$ and so
\begin{equation}
\frac{\partial}{\partial\beta}\left(\ln(Z)\right) =
\langle \ln(P(\data| \params))\rangle_\beta.
\end{equation}
Using the likelihoods calculated in the MCMC simulations, we numerically
integrate the averaged log-likelihood over $\beta$ which yields an estimation
of the marginal likelihood. To be confident that the estimate is correct, we
ensure that we use a sufficient number of temperatures and that they cover the
region of interest.

\subsection{Results}
Applying the thermodynamic integration technique to all the models, we estimate
the evidence for each model. Taking the ratio of the evidences gives us the
Bayes factor and since we set the ratio of the prior on the models to unity,
the Bayes factor is exactly the odds ratio (see Eqn.~\eqref{eqn: odds-ratio}).

We present the $\log_{10}\textrm{odds-ratio}$ between the models in
Table~\ref{tab: log odds-ratio}. A positive value indicates
that the data prefers model A over model B.
Note that the error here is an estimate of the systematic error due to the
choice of $\beta$ values \citep[see][for details]{Foreman-Mackay2013}.
\begin{table}
\caption{Tabulated log-odds-ratios for all models. $^{*}$By the precession
model here we mean the precession with a modified Gaussian beam model as
discussed in Sec.~\ref{sec: refining the beam-width model}.}
\label{tab: log odds-ratio}
\begin{tabular}{ccc}\hhline{===}
Model A & Model B & $\log_{10}(\textrm{odds-ratio})$ \\ \hline
switching & noise-only &
$\oddsBeamwidthSwitchingBeamwidthNoiseOnly \pm \errBeamwidthSwitchingBeamwidthNoiseOnly$ \\
precession$^{*}$ & noise-only &
$\oddsBeamwidthModifiedGaussianBeamwidthNoiseOnly \pm \errBeamwidthModifiedGaussianBeamwidthNoiseOnly $ \\
precession$^{*}$ & switching &
$\oddsBeamwidthModifiedGaussianBeamwidthSwitching \pm \errBeamwidthModifiedGaussianBeamwidthSwitching $ \\
\hhline{===}
\end{tabular}
\end{table}

This table allows quantitative discrimination amongst the models. The first two
rows compare the switching and modified-Gaussian precession models against the
noise-only model with the periodic modulating models being strongly preferred
in both cases. Then in the last row we present the log-odds-ratio between the
modified-Gaussian precession and switching model which shows that the data
prefers the precession Modified Gaussian model by a factor
$10^{\oddsBeamwidthModifiedGaussianBeamwidthSwitching}$. Using the interpretation
of \mboxcitet{jeffreys1998theory}, the strength of this evidence can be interpreted
as `decisive' in favour of this precession model. For completeness, we also
mention that the odds-ratio for the non-modified Gaussian model (which failed
to fit the data in a physically meaningful way) against the noise-only model
was $\oddsBeamwidthGaussianBeamwidthNoiseOnly \pm
\errBeamwidthGaussianBeamwidthNoiseOnly$.

\subsection{Effect of the choice of prior}
\label{sec: effect of the choice of prior}

For both beam-widths in the switching model we used uniform priors on $[0,
f\PATNF]$ with $f=0.1$ and these were transformed to also provide a fair prior
on $\rho_2^0$ and $\rho_2''$. This choice of $f$ was taken from the upper limit
quoted in \citet{Lyne1988} for typical values of the pulse width.
Nevertheless, changing $f$ can have a measurable impact on the odds-ratio and
so we will now study this effect.

To begin, we rewrite
Eqn.~\eqref{eqn: marginal likelihood}, the marginal
likelihood, by factoring out the $N$ parameters which have a uniform prior
\begin{align}
P(\data| \M_i) =
\frac{1}{\prod_{i}^{N}(b_i - a_i)}
\int P(\data| \params, \M_i)P(\params^{*}|\M_{i}) d\params,
\end{align}
where by $P(\params^{*}| \M_i$) we mean the probability distribution of all
remaining parameters which are not factored out, and $[a_i, b_i]$ is the range
for the $i^{th}$ uniform parameter.  For the switching beam-width, the prefactor of
this integral (factoring out the prior on $\Wone$ and $\Wtwo$) is
$(f\PATNF)^{-2}$: varying $f$ directly impacts the evidence for the switching
model. For the precession model, we cannot factor the dependence on $f$ in the
same way as we use a central normal prior on $\rho_2''$.  We set the
standard-deviation of this prior by applying Eqn.~\eqref{eqn: rho2dd prior} so
that it is inversely proportional to $f$. If both the prior on $\rho_2^0$ and
$\rho_2''$ had been proportional to $f$ we would have an exact cancellation in
the odds-ratio and hence no dependence on $f$. This is not the case and due to
our prior on $\rho_2''$ the odds-ratio will depend on $f$.

\begin{figure}
\centering
\includegraphics[width=0.45\textwidth]{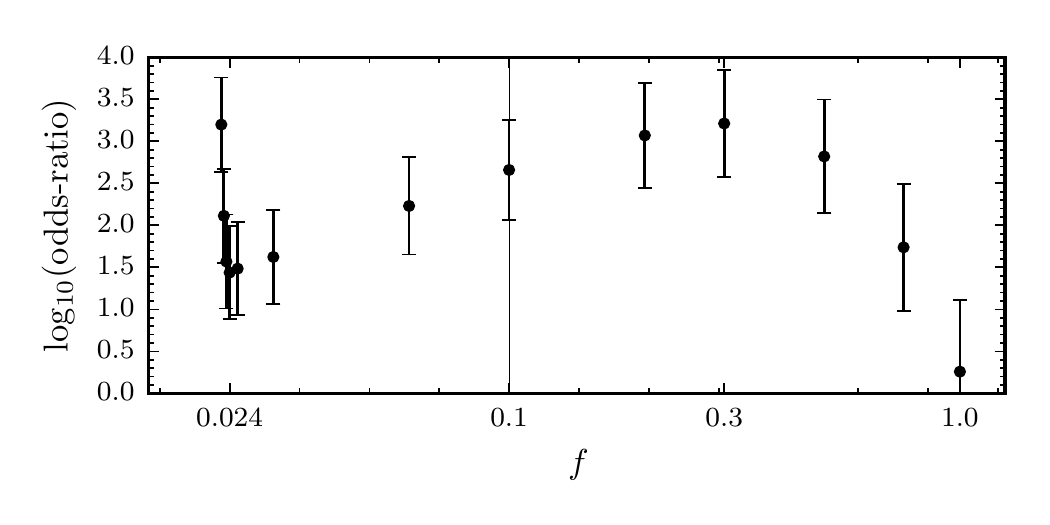}
\caption{Dependence of the
         odds-ratio with $f$, the fraction of $\PATNF$ used to constrain the
         beam-width priors. The vertical line
         marks the choice of $f=0.1$ used in our model comparison based on the
         upper limit given by \citet{Lyne1988}}
\label{fig: beamwidth switching frac variation}
\end{figure}

To test the dependence, in Fig.~\ref{fig: beamwidth switching frac variation}
we plot the log odds-ratio as a function of $\log_{10}(f)$ (note that $f=0.1$
corresponds to the result in Table.~\ref{tab: log odds-ratio}). There are
several features to understand. First, for $f\lesssim 0.024$ the
odds-ratio rapidly grows, favouring precession; this is because for such small
values of $f$ the beam-width switching prior excludes the values of
$\Wone$ required to fit the data. As a result the switching solutions are
unnaturally disadvantaged compared to the precession solutions, such odds-ratios do not
fairly compare the models.

For $0.024 \lesssim f \lesssim 0.3$ the log-odds-ratio is approximately linear
growing from $1.44$ when $f=0.24$ to $3.21$ when $f=0.3$. In this region the
solutions for both models are supported by the prior in that it does not
exclude or disfavour the posterior value. The variation in the odds-ratio
results from changes in the prior volume of the switching model, the evidence
for the precession model is constant in this region.  Small $f$ values
maximally constrains the prior volume for the switching model (without
excluding posterior values) and hence give the greatest weight of evidence to
switching and the smallest odds-ratios. For larger $f$ values the log of the
prior volume grows linearly resulting in the observed growth.

For $f \gtrsim 0.3$ our choice of standard-deviation for $\rho_2''$ starts to
disfavour the posterior value because it is inversely proportional to $f$. As
a result the evidence for the precession model decreases faster than the loss of
evidence for the switching model leading to the observed drop in the odds-ratio.
In this case it is the precession solutions which are unnaturally disadvantaged
by our choice of prior and so, as in the $f\lesssim 0.024$ case, we do not consider
such odds-ratios as a fair comparison of the models.

In summary the log-odds-ratio and hence our conclusion is robust to reasonable
variations in $f$ from 0.05 to 0.5.

\section{Discussion}
\label{sec: discussion}

In this work we are using a data set (provided by \mboxcitet{Lyne2010}) on the
spin-down and beam-width of B1828-11 to compare models for the observed
periodic variations.  The two concepts under consideration are free precession
and magnetospheric switching.  In order to be quantitative, we built signal
models for the beam-width and spin-down from these conceptual ideas. Using the
spin-down data to create proper, physically motivated priors for the beam-width
parameters, we then perform a Bayesian model comparison between the models
asking ``which model does the beam-width data support?". For the models
considered here, the data most strongly supports a precession model with a
modified-Gaussian beam geometry allowing for an elliptical beam where the
ellipticity has a latitudinal dependence.

To be clear, this does not rule out the switching interpretation since we have
not tested an exhaustive set of models---we can only compare between particular
models. As an example we could imagine modifying the switching model such that
either the switching times, or the magnetospheric states are probabilistic (or
a combination of the two).  Further we believe there is good grounds to develop
models combining the precession and switching interpretation like those
discussed in \mboxcitet{Jones2012}.

In addition to the data considered in this work, a number of
high-time-resolution observations of B1828-11 were performed
by the Parkes telescope, as discussed in \citet{Stairs2003}.
This data set shows interesting variability in beam width on short
timescales of ~O(100) pulses. While the qualitative "noisiness" of the
beam-width data is already apparent from the current data-set (eg. see
Fig.1), such high-time-resolution data could be very interesting to
include in a more detailed future model comparison.

The process of fitting the models to the data and performing posterior
predictive checks also provides a mechanism to evaluate the models. For both
spin-down models the maximum posterior plots with the data (Figs.~\ref{fig:
perera spin-down posterior}B and \ref{fig: precession spin-down posterior}B)
revealed a systematic failure to fit the second (slightly lower) minima. This
suggests new ingredients could be introduced to both models to explain this.

The posteriors for the precession model indicate that B1828-11 is a
near-orthogonal rotator and we observe it from close to the equatorial plane.
If this is the case, and the two beams of the pulsar are symmetric about the
origin, then we expect to see the second beam as an interpulse. Indeed, we
discuss further in appendix~\ref{sec: implications for the unobserved beam} how
during the precessional cycle we should expect the intensity of this second
beam to dominate at certain phases. Since no such interpulse is reported,
either the second beam is weaker, or the beams must have a kink of greater than
$4.6\degr$ (see appendix~\ref{sec: implications for the unobserved beam}).

In this work we have developed the framework to evaluate models for the
variations observed in B1828-11. This is not intended as an exhaustive review
of all models, but rather a discussion on the intricacies that arise such as
setting up proper and well-motivated priors. This work lays the groundwork
for a more exhaustive test of all available models and can also be extended
by improvements to the data sources.

\section*{Acknowledgements}
GA acknowledges financial support from the University of Southampton and the
Albert Einstein Institute (Hannover).  DIJ acknowledge support from STFC via
grant number ST/H002359/1, and also travel support from NewCompStar (a
COST-funded Research Networking Programme).  We especially thank Will Farr,
Danai Antonopoulou, and Ben Stappers for valuable discussions and comments,
\mboxcitet{dan_foreman_mackey_2014_11020} for the software used in generating
posterior probability distributions, and \mboxcitet{Lyne2010} for generously sharing
the data for B1828-11.

\bibliographystyle{mnras}
\bibliography{bibliography}

\appendix

\section{Procedure for MCMC parameter estimation}
\label{sec: procedure for the mcmc parameter estimation}

The procedure used to simulate the posterior distribution can determine the
quality of the estimation. Therefore we will now set out an algorithmic method
to ensure the our results are reproducible.

To estimate the posterior given a signal model and prior, we run two MCMC
simulations: an initialisation and production. In the following the term
\emph{walker} refers to single chains in the MCMC simulation, the
\mboxcitet{Foreman-Mackay2013} implementation runs a number of these in parallel
for each simulation.

\begin{itemize}

\item For the initialisation run, we draw samples from the prior distribution
to set the initial parameters for each walker. The simulation therefore has the
chance to explore the entire parameter space. After a sufficient number of
steps, the walkers will converge to the local maxima in the log-likelihood. By
visually inspecting the data we determine the nature of the local maxima: in
all cases a single maxima dominated such that, given a sufficient length of
simulation, we expect all walkers to converge to this maxima. Alternatively we
could have found multiple similarly strong maxima, in this case further
analysis would be required. This was not found to be the case for any of the
models in this analysis.

\item For the second step we set the initial state of 100 walkers by uniformly
dispersing them in a small range about the maximum-likelihood found in the
previous step. The simulation proceeds from this initial state and we divide
the resulting samples equally into two: discarding the first half as a
so-called `burn-in', we retain the second half as the production data used to
estimate the posterior.  The burn-in removes any memory of the artificial
initialisation of the walkers at the start of this step.

\end{itemize}

Having run an MCMC simulation we check that the chains have properly converged
(for a discussion on this see \mboxcitet{gelman2013bayesian}).  The MCMC
simulations provide an estimate of the posterior densities for the model
parameters. We will also perform `posterior predictive checks' to ensure the
posterior is a suitable fit to the data, i.e. we compare the data to the model
prediction when the model parameters are set to the values corresponding to the
peaks of the posterior probability  distributions.

\section{Implications for the unobserved beam}
\label{sec: implications for the unobserved beam}
The precession model developed here assumes the observer only ever sees one
pole of the beam-axis, but in the canonical model we often imagine there is
also emission from an opposite magnetic pole. In several pulsars this can be
seen as an \emph{interpulse} $180\degr$ out of phase from the main pulse
\citep{Lyne1988, Maciesiak2011}; these pulsars are generally found to be close
to orthogonal rotators\footnote{The use of `interpulse' here strictly refers to
seeing the opposite beam of the pulsar, and not cases where the pulsar is
almost aligned and interpulses are thought to come from the same beam}. No
such interpulse is reported for B1828-11, we will now discuss the implications
of this given the precession interpretation.

Let us imagine a scenario where the observer is in the northern (magnetic)
hemisphere and label the beam protruding into their hemisphere (which they will
see with the greater intensity due to the smaller angular separation) at the
start of the thought experiment as the north pole.  Then, when $\Theta < \pi/2$,
the north and south
pole make angles $\Theta$ and $\pi-\Theta$ respectively with the fixed angular
momentum vector. Now we see that if, during the course of the precessional
cycle, $\Theta > \pi/2$ then the south pole will protrude into the northern
hemisphere and the north pole into the southern hemisphere. Provided both poles
are identical, but regardless of the details of the beam-geometry, at this time
we must expect the observer to see the south pole at greater intensity than the
north pole. An example of this is shown in Fig.~\ref{fig: J observer plane
other pole}, but note that the observer will see the greatest intensity from
the south-pole half a rotation after this instance.

Our posterior distributions inform us that, if the precession interpretation is
correct, we are in exactly this situation: $\Theta$ ranges from $85.8\degr$
to $92.3\degr$ over a precessional cycle\footnote{Numbers generated from the
maximum posterior estimates of $\chi$ and $\theta$ using the spin-down data.
The estimated error for both values is $\pm0.08\degr$} so we should see the
interpulse.

This is readily explained if the south pole is substantially weaker in
intensity, or by the one-pole interpretation \mboxcitet{Manchester1977}.
Alternatively, it could be that  the two beams are not diametrically opposed
but are latitudinally `kinked'. In the later case we can put a lower bound on
the kink angle by requiring that the polar angle of the south pole is always
greater than that of the north (see Fig.~\ref{fig: J observer  plane other
pole}). From our MPE this gives a lower bound of $4.6\degr$ for the polar kink
angle. This latitudinal kink can be compared with the longitudinal kink of
interpulses observed in other pulsars: often these are not found at exactly
$180\degr$, but can deviate by 10's of degrees (see the separation of
interpulses for double-pole interpulses in Table~1. of \cite{Maciesiak2011}).
Allowing for such kinks in both beams is a possible extension to the
precessional model.

\begin{figure}
\centering
\includegraphics[width=0.4\textwidth]{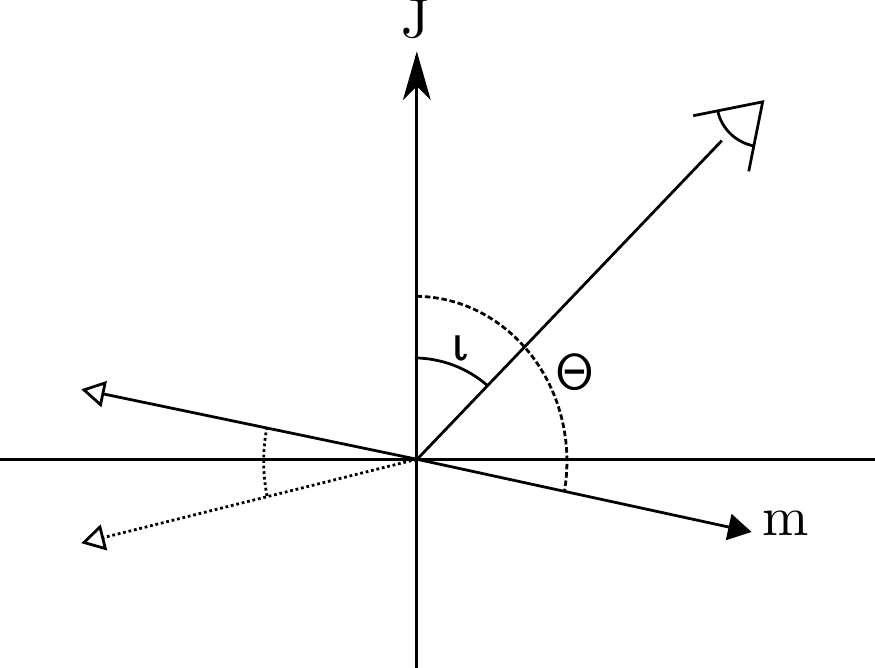}
\caption{Extension to Fig.~\ref{fig: J observer plane} adding in the south pole
(demarked by a white triangle and solid line) for an instance in which the
north pole is in the southern hemisphere. The dotted line with a white
arrowhead is the `kinked' beam.  For clarity, the north and south poles are
shown at times that differ by half a rotation period.}
\label{fig: J observer plane other pole}
\end{figure}

\bsp
\label{lastpage}
\end{document}

%% file: Evidence.tex
\def\oddsBeamwidthModifiedGaussianBeamwidthNoiseOnly{60.1}
\def\errBeamwidthModifiedGaussianBeamwidthNoiseOnly{0.5}
\def\oddsBeamwidthSwitchingBeamwidthNoiseOnly{57.4}
\def\errBeamwidthSwitchingBeamwidthNoiseOnly{0.5}
\def\oddsBeamwidthGaussianBeamwidthNoiseOnly{3.1}
\def\errBeamwidthGaussianBeamwidthNoiseOnly{0.6}
\def\oddsBeamwidthModifiedGaussianBeamwidthSwitching{2.7}
\def\errBeamwidthModifiedGaussianBeamwidthSwitching{0.5}

%% file: Table_Spindown_Switching_prior.tex
\begin{tabular}{lll} \hhline{===}
        Parameter & Distribution &  Units\\ \hline
$\dot{\nu}_{1}$ & Unif(${\text{\textrm{-}}3.66}\times 10^{\textrm{-}13}$, ${\text{\textrm{-}}3.64}\times 10^{\textrm{-}13}$) & $\mathrm{s}^{\textrm{-}2}$\\
$\dot{\nu}_{2}$ & Unif(${\text{\textrm{-}}3.67}\times 10^{\textrm{-}13}$, ${\text{\textrm{-}}3.66}\times 10^{\textrm{-}13}$) & $\mathrm{s}^{\textrm{-}2}$\\
$\ddot{\nu}$\textrm{-}$\ddot{\nu}^{\tiny\mathrm{ATNF}}$
 & $\mathcal{N}$(0, ${9.0}\times 10^{\textrm{-}27}$) & $\mathrm{s}^{\textrm{-}3}$\\
$T$ & Unif(450, 550) & days\\
$t_\mathrm{A}$ & Unif(0, 250) & days\\
$t_\mathrm{B}$ & Unif(0, 250) & days\\
$t_\mathrm{C}$ & Unif(0, 250) & days\\
$\phi_{0}$ & Unif(0, 1) & \\
$\sigma_{\dot{\nu}}$ & Unif(0, ${1}\times 10^{\textrm{-}15}$) & $\mathrm{s}^{\textrm{-}2}$\\
\hhline{===}
\end{tabular}

%% file: Table_Spindown_Switching_posterior.tex
\begin{tabular}{lll} \hhline{===}
        Parameter & Mean $\pm$ s.d. &  Units\\ \hline
$\dot{\nu}_{1}$ & ${\text{\textrm{-}}3.6489}\times 10^{\textrm{-}13}$$\pm$${6.33}\times 10^{\textrm{-}17}$ & $\mathrm{s}^{\textrm{-}2}$\\
$\dot{\nu}_{2}$ & ${\text{\textrm{-}}3.6635}\times 10^{\textrm{-}13}$$\pm$${4.44}\times 10^{\textrm{-}17}$ & $\mathrm{s}^{\textrm{-}2}$\\
$\ddot{\nu}$\textrm{-}$\ddot{\nu}^{\tiny\mathrm{ATNF}}$
 & ${\text{\textrm{-}}3.1051}\times 10^{\textrm{-}28}$$\pm$${9.0}\times 10^{\textrm{-}27}$ & $\mathrm{s}^{\textrm{-}3}$\\
$T$ & $485.52$$\pm$0.8649 & days\\
$t_\mathrm{A}$ & $157.75$$\pm$7.6587 & days\\
$t_\mathrm{B}$ & $159.71$$\pm$11.7798 & days\\
$t_\mathrm{C}$ & 15.1379$\pm$4.3925 & days\\
$\phi_{0}$ & 0.5278$\pm$0.0143 & \\
$\sigma_{\dot{\nu}}$ & ${4.0932}\times 10^{\textrm{-}16}$$\pm$${1.84}\times 10^{\textrm{-}17}$ & $\mathrm{s}^{\textrm{-}2}$\\
\hhline{===}
\end{tabular}

%% file: Table_Beamwidth_Switching_0.1_prior.tex
\begin{tabular}{lll} \hhline{===}
        Parameter & Distribution &  Units\\ \hline
$W_{1}$$^{}$ & Unif(0, 40.5000) & ms\\
$W_{2}$$^{}$ & Unif(0, 40.5000) & ms\\
$T$$^{*}$ & $\mathcal{N}$($485.5$, 0.8649) & days\\
$t_\mathrm{A}$$^{*}$ & $\mathcal{N}$($158.0$, 7.6587) & days\\
$t_\mathrm{B}$$^{*}$ & $\mathcal{N}$($160.0$, 11.7798) & days\\
$t_\mathrm{C}$$^{*}$ & $\mathcal{N}$(15.1379, 4.3925) & days\\
$\phi_{0}$$^{*}$ & $\mathcal{N}$(0.5278, 0.0143) & \\
$\sigma_{W_{10}}$$^{}$ & Unif(0, 5) & ms\\
\hhline{===}
\end{tabular}

%% file: Table_Beamwidth_Switching_0.1_posterior.tex
\begin{tabular}{lll} \hhline{===}
        Parameter & Mean $\pm$ s.d. &  Units\\ \hline
$W_{1}$ & 9.5166$\pm$0.0956 & ms\\
$W_{2}$ & 7.2327$\pm$0.0830 & ms\\
$T$ & $485.04$$\pm$0.7286 & days\\
$t_\mathrm{A}$ & $150.98$$\pm$4.3909 & days\\
$t_\mathrm{B}$ & $155.29$$\pm$3.0598 & days\\
$t_\mathrm{C}$ & 16.4134$\pm$4.5771 & days\\
$\phi_{0}$ & 0.5409$\pm$${8.38}\times 10^{\textrm{-}3}$ & \\
$\sigma_{W_{10}}$ & 1.5964$\pm$0.0427 & ms\\
\hhline{===}
\end{tabular}

%% file: Table_Spindown_Precession_prior.tex
\begin{tabular}{lll} \hhline{===}
        Parameter & Distribution &  Units\\ \hline
$\tau_{\mathrm{Age}}$\textrm{-}$\tau_{\mathrm{Age}}^{\tiny\textrm{ATNF}}$
 & $\mathcal{N}$(0, 0.3169) & yrs\\
$n$\textrm{-}$n^{\tiny\textrm{ATNF}}$ & $\mathcal{N}$(0, 0.1700) & \\
$P$\textrm{-}$P^{\tiny\textrm{ATNF}}$
 & $\mathcal{N}$(0, ${1.2}\times 10^{\textrm{-}11}$) & s\\
$\tau_{P}$ & Unif(450, 550) & days\\
$\theta$ & Unif(0, 0.1) & rad\\
$\chi$ & Unif($2\pi/5$, $\pi/2$) & rad\\
$\psi_0$ & Unif(0, $2\pi$) & rad\\
$\sigma_{\dot{\nu}}$ & Unif(0, ${1}\times 10^{\textrm{-}15}$) & $\mathrm{s}^{\textrm{-}2}$\\
\hhline{===}
\end{tabular}

%% file: Table_Spindown_Precession_posterior.tex
\begin{tabular}{lll} \hhline{===}
        Parameter & Mean $\pm$ s.d. &  Units\\ \hline
$\tau_{\mathrm{Age}}$\textrm{-}$\tau_{\mathrm{Age}}^{\tiny\textrm{ATNF}}$
 & ${7.461}\times 10^{\textrm{-}3}$$\pm$0.3159 & yrs\\
$n$\textrm{-}$n^{\tiny\textrm{ATNF}}$ & 0.0199$\pm$0.1701 & \\
$P$\textrm{-}$P^{\tiny\textrm{ATNF}}$
 & ${1.0436}\times 10^{\textrm{-}14}$$\pm$${1.19}\times 10^{\textrm{-}11}$ & s\\
$\tau_{P}$ & $485.56$$\pm$0.8188 & days\\
$\theta$ & 0.0490$\pm$0.0020 & rad\\
$\chi$ & 1.5517$\pm$0.0013 & rad\\
$\psi_0$ & 3.8709$\pm$0.0697 & rad\\
$\sigma_{\dot{\nu}}$ & ${4.0423}\times 10^{\textrm{-}16}$$\pm$${1.81}\times 10^{\textrm{-}17}$ & $\mathrm{s}^{\textrm{-}2}$\\
\hhline{===}
\end{tabular}

%% file: Table_Beamwidth_Gaussian_prior.tex
\begin{tabular}{lll} \hhline{===}
        Parameter & Distribution &  Units\\ \hline
$\tau_{P}$$^{*}$ & $\mathcal{N}$($485.6$, 0.8188) & days\\
$P$\textrm{-}$P^{\tiny\textrm{ATNF}}$
$^{*}$ & $\mathcal{N}$(${1.04}\times 10^{\textrm{-}14}$, ${1.19}\times 10^{\textrm{-}11}$) & s\\
$\theta$$^{*}$ & $\mathcal{N}$(0.0490, 0.0020) & rad\\
$\chi$$^{*}$ & $\mathcal{N}$(1.5517, 0.0013) & rad\\
$\psi_0$$^{*}$ & $\mathcal{N}$(3.8709, 0.0697) & rad\\
$\rho$$^{}$ & Unif(0, 0.1500) & rad\\
$\cos(\iota)$$^{}$ & Unif(\textrm{-}1, 1) & \\
$\sigma_{W_{10}}$$^{}$ & Unif(0, 5) & ms\\
\hhline{===}
\end{tabular}

%% file: Table_Beamwidth_ModifiedGaussian_0.1_prior.tex
\begin{tabular}{lll} \hhline{===}
        Parameter & Distribution &  Units\\ \hline
$\tau_{P}$$^{*}$ & $\mathcal{N}$($485.6$, 0.8188) & days\\
$P$\textrm{-}$P^{\tiny\textrm{ATNF}}$
$^{*}$ & $\mathcal{N}$(${1.04}\times 10^{\textrm{-}14}$, ${1.19}\times 10^{\textrm{-}11}$) & s\\
$\theta$$^{*}$ & $\mathcal{N}$(0.0490, 0.0020) & rad\\
$\chi$$^{*}$ & $\mathcal{N}$(1.5517, 0.0013) & rad\\
$\psi_0$$^{*}$ & $\mathcal{N}$(3.8709, 0.0697) & rad\\
$\rho_{2}^{0}$$^{}$ & Unif(0, 0.1464) & rad\\
$\rho_{2}''$$^{}$ & $\mathcal{N}$(0, 6.8308) & rad$^{\textrm{-}2}$\\
$\cos(\iota)$$^{}$ & Unif(\textrm{-}1, 1) & \\
$\sigma_{W_{10}}$$^{}$ & Unif(0, 5) & ms\\
\hhline{===}
\end{tabular}

%% file: Table_Beamwidth_ModifiedGaussian_0.1_posterior.tex
\begin{tabular}{lll} \hhline{===}
        Parameter & Mean $\pm$ s.d. &  Units\\ \hline
$\tau_{P}$ & $484.87$$\pm$0.4706 & days\\
$P$\textrm{-}$P^{\tiny\textrm{ATNF}}$
 & ${\text{\textrm{-}}1.7719}\times 10^{\textrm{-}13}$$\pm$${1.19}\times 10^{\textrm{-}11}$ & s\\
$\theta$ & 0.0490$\pm$0.0020 & rad\\
$\chi$ & 1.5517$\pm$0.0013 & rad\\
$\psi_0$ & 3.9701$\pm$0.0403 & rad\\
$\rho_{2}^{0}$ & 0.0245$\pm$0.0004 & rad\\
$\rho_{2}''$ & 3.4421$\pm$0.3878 & rad$^{\textrm{-}2}$\\
$\cos(\iota)$ & ${7.9326}\times 10^{\textrm{-}3}$$\pm$${1.9}\times 10^{\textrm{-}3}$ & \\
$\sigma_{W_{10}}$ & 1.5833$\pm$0.0422 & ms\\
\hhline{===}
\end{tabular}